\documentclass{getwriting}
\usepackage{amsmath}
\usepackage{bm}
\usepackage{booktabs}

\usepackage[labelfont=bf]{caption}
\title{Bayesian Decision Curve Analysis with bayesDCA}
\author[1,2]{Giuliano N.F. Cruz \orcidlink{0000-0001-9122-8286}}
\author[1,2,3]{Keegan Korthauer \orcidlink{0000-0002-4565-1654} \thanks{keegan@stat.ubc.ca}}

\affil[1]{\footnotesize Faculty of Science, University of British Columbia}
\affil[2]{\footnotesize British Columbia Children’s Hospital Research Institute}
\affil[3]{\footnotesize Department of Statistics, University of British Columbia}
\begin{document}
\maketitle
\begin{abstract}

\noindent Clinical decisions are often guided by clinical prediction models or diagnostic tests. Decision curve analysis (DCA) combines classical assessment of predictive performance with the consequences of using these strategies for clinical decision-making. In DCA, the best decision strategy is the one that maximizes the so-called net benefit: the net number of true positives (or negatives) provided by a given strategy. In this decision-analytic approach, often only point estimates are published. If uncertainty is reported, a risk-neutral interpretation is recommended: it motivates further research without changing the conclusions based on currently-available data. However, when it comes to new decision strategies, replacing the current Standard of Care must be carefully considered -- prematurely implementing a suboptimal strategy poses potentially irrecoverable costs. In this risk-averse setting, quantifying uncertainty may also inform whether the available data provides enough evidence to change current clinical practice. Here, we employ Bayesian approaches to DCA addressing four fundamental concerns when evaluating clinical decision strategies: (i) which strategies are clinically useful, (ii) what is the best available decision strategy, (iii) pairwise comparisons between strategies, and (iv) the expected net benefit loss associated with the current level of uncertainty. While often consistent with frequentist point estimates, fully Bayesian DCA allows for an intuitive probabilistic interpretation framework and the incorporation of prior evidence. We evaluate the methods using simulation and provide a comprehensive case study. Software implementation is available in the bayesDCA R package. Ultimately, the Bayesian DCA workflow may help clinicians and health policymakers adopt better-informed decisions.
\\[10pt] 

\end{abstract}
\newpage
\section{Introduction}

In Decision Curve Analysis (DCA), we are typically interested in estimating the net benefit of adopting a given clinical decision strategy \cite{Vickers_2006}.
A decision strategy may be simply treating all patients under suspicion of a given condition because the condition is
so deadly that any costs of a potentially unnecessary intervention are outweighed by devastating costs of neglecting necessary treatment, even if the risk is small -- e.g., treating potentially aggressive cancer.
In this ``Treat all" strategy, intervention happens in all patients regardless of the true underlying disease status -- e.g., whether the patient's cancer is aggressive or not. Conversely, an intervention may be high risk while the underlying condition is benign -- e.g., surgical removal of a stable noncancerous brain tumour.
In this case, a reasonable decision strategy is to not treat any patient -- the ``Treat none" strategy.

In general, the relative risk conferred by the disease and the treatment does not always clearly side with either the ``Treat all" or ``Treat none" strategies. In addition, there is often uncertainty around a patient's disease status or prognosis. In this case, a decision strategy could be based on a predictive model that estimates, e.g., a patient's likelihood of having aggressive disease. If the patient's likelihood
is above a decision threshold $t$, then we intervene. Beyond the probability of having a disease right now (diagnostic setting), the threshold $t$ could also be the probability of a future event like death, hospitalization, or disease progression (prognostic setting). The same idea serves for binary tests, in which case we intervene if the test is positive. 

In the context of DCA, the Net Benefit (NB) at the decision threshold $t$ can be written as\cite{Vickers_2006}: 
\begin{align}
    NB_t = \Big(TP_t - FP_t \cdot w_t \Big)/n\label{nb}
\end{align}
where $TP_t$ and $FP_t$ are corresponding true and false positive counts, $n$ is the total sample size, and $w_t = t/(1-t)$. Given a decision threshold $t$, this definition (\ref{nb}) fixes the weight of each true positive at 1, which mathematically implies a relative weight of $w_t$ for each false positive. This allows decision analysis without the need to specify the absolute costs of each potential outcome (true and false positive/negative). Instead, we rely on a clinically-motivated decision threshold $t$ which properly weights true and false positives/negatives based on the clinical context\cite{Wynants_2019}. To consider a range of relative weights (e.g., due to disagreement between clinicians or even patients' preferences), the decision curve is then constructed by plotting $NB_t$ for a reasonable range of decision thresholds $t$.

At each decision threshold, the optimal decision strategy is the one that maximizes the expected net benefit\cite{Vickers_2006}. If using one strategy imposes a higher cost than using another, then its net benefit may be adjusted by subtracting a ``test harm" term which represents the strategy-specific costs \cite{Vickers_2019}. However, this approach requires an additional, subjective step of calculating the cost for each decision strategy under investigation. Moreover, strategy costs may be highly context-specific. For example, they may depend on resources available in a given location. This does not undermine the value of calculating the likely costs of each strategy, but it acknowledges that the challenges of this task may exceed the scope of model development and validation studies. 

When interpreting DCA results, there may be special considerations for making a decision that changes a well-accepted practice. Beyond the one-time cost of the implementation process itself, there is always the risk of implementing the ``wrong" strategy whose apparent optimality was an artifact of chance. As with any other estimate, the observed net benefit is subject to random variation in the data. Addressing this uncertainty, however, depends on how we understand risk. Under risk neutrality, we only care about expected gains (or costs) and uncertainty quantification does not change which strategy should be used, although it can motivate further research\cite{McKenna_2015, Vickers_2019, Glynn_2023}. On the other hand, prematurely replacing the current Standard of Care (SoC) with a new decision strategy poses potentially irrecoverable costs to individual patients, healthcare institutions, and even healthcare systems\cite{Vickers_2008, Glynn_2023}. This setting motivates risk aversion, which may require a more careful assessment of uncertainty to prevent premature implementation.

The work presented here regards the Bayesian estimation of net benefit and is a priori indifferent as to whether end users are risk-neutral or not -- a debate we do not aim to settle. To describe the full potential of the method, we include a case study that does not assume that risk neutrality is satisfied when assessing a new decision strategy to potentially replace the SoC. If we are too uncertain about the net benefit of each decision strategy under investigation, or if the net benefit gain from adopting a new decision strategy is negative with high probability, then more data may be desirable before changing current practice. Given the superiority of a decision strategy in terms of observed net benefit, different levels of uncertainty may be compatible with context-specific costs to ultimately justify implementation. Still, the risk-neutral reader is free to interpret uncertainty as a motivation for further research only, without changing the conclusions based on the observed estimates, if desired.

Therefore, uncertainty quantification around estimates of net benefit and their differences allows two potential interpretations, depending on the reader's risk profile. Under risk neutrality, uncertainty informs whether more research is needed but does not change our decision based on the currently-available point estimates. Under risk aversion, uncertainty may inform whether the available data provides enough evidence to change well-established clinical practice. Both interpretations of uncertainty in DCA motivate further research, but only the risk-averse wait for more data before changing clinical practice currently in place. In what follows, we employ Bayesian approaches to DCA addressing four fundamental concerns when evaluating clinical decision strategies:
\begin{enumerate}[topsep=0pt, partopsep=0pt]
    \itemsep0em 
    \item Which strategies are clinically useful?
    \item What is the best available decision strategy?
    \item Direct pairwise comparisons between strategies.
    \item What is the expected net benefit loss associated with the current level of uncertainty?
\end{enumerate}
Uncertainty around these concerns is natural to address in the Bayesian context, however the approach remains agnostic to its interpretation. While often consistent with the frequentist approach in terms of point estimates (e.g., under vague priors), the fully Bayesian estimation of net benefit allows for an intuitive probabilistic interpretation of DCA results as well as for the principled incorporation of prior evidence. Our proposal builds on the work from Wynants et al. (2018) \cite{Wynants_2018} which proposed Bayesian DCA for evidence synthesis in meta-analysis of data from multiple settings (e.g., multiple hospitals). We adapt their binary outcome model to allow for the incorporation of prior information in the single-setting case and propose an alternative formulation for survival outcomes. We then compare the methodology with Frequentist alternatives using simulation and provide a case study with openly available data. The proposed approaches are implemented in the freely available bayesDCA R package.

\section{Results}
\subsection{General setting}
Suppose we have access to validation data to assess one or more clinical decision strategies, be it a pre-specified clinical prediction model or a binary diagnostic/prognostic test. We will use DCA to decide whether any of the strategies under investigation is clinically useful and which of them is the best. We will also examine the difference between strategies, which may be useful to evaluate alternatives under scenarios where one or more strategies is unavailable. Finally, we would like to have a sense of whether the current study is precise enough so that new studies with the same population are not necessary. In what follows, we first describe the Bayesian estimation of decision curves for binary outcomes and then extend it to survival outcomes. 

\subsection{Bayesian DCA for binary outcomes}
The Net Benefit formulation in (\ref{nb}) can be rewritten in terms of the outcome prevalence ($p$) and the threshold-specific Sensitivity (Se) and Specificity (Sp).
\begin{align}
    \textrm{NB}_{t} = \textrm{Se}_{t}\cdot p - (1 - \textrm{Sp}_{t})\cdot(1-p)\cdot w_t \label{nb-p=se-sp}
\end{align}
During (external) validation of predictive models or binary tests, we can estimate the parameters above using a conjugate Beta-Bernoulli joint model for the indicator variables of positive predictions and disease status (see section \textit{\nameref{sec:bdca-details}} for full model specification). Given conjugacy, the full posterior distribution of the parameter vector $\left(p,\, \textrm{Se}_{t},\, \textrm{Sp}_{t}\right)$ is known in closed form:
\begin{align}
\textrm{Beta}(p \vert D+\alpha_0, ND+\beta_0)
\ \times \ 
\textrm{Beta}(\textrm{Se}_t \vert TP_t+\alpha_1, FN_t+\beta_1) 
\ \times \ 
\textrm{Beta}(\textrm{Sp}_t \vert TN_t+\alpha_2, FP_t+\beta_2)
\label{bayes-dca-model}
\end{align}
where $TP_t,\ FP_t,\ TN_t,\ FN_t,\ D,\textrm{ and } ND$ represent the total number of true and false positives, true and false negatives, and individuals with and without the disease, respectively. The  $\big(\alpha_\cdot, \beta_\cdot\big)$ terms are parameters of the independent Beta prior distributions. Within bayesDCA, we set $\alpha_\cdot=\beta_\cdot=1$ as a default, representing uniform priors on the $(0, 1)$ interval, though the user may choose different priors as well. We suggest a more informative prior based on the expected relationship between the decision threshold and sensitivity/specificity in the Supplement (section \textit{\nameref{supp:inf-priors}} and Supplementary Figures (\ref{fig:ppc}--\ref{fig:ppc2})). In addition to conjugacy, the factorization in (\ref{bayes-dca-model}) is due to parameter orthogonality in the likelihood function and implies posterior independence. This means that Markov-Chain Monte Carlo (MCMC) is not needed: we can combine samples from the marginal posteriors to easily generate valid samples from the joint posterior, making estimation particularly fast - typically a fraction of a second for an entire DCA.

The model (\ref{bayes-dca-model}) can be seen as a single-setting version of the model from Wynants et al. (2018)\cite{Wynants_2018}, which models the parameters of interest on the logit scale using a multivariate normal (MVN) distribution. Given the meta-analysis context, there are multiple sensitivities, specificities, and prevalences to be considered at each threshold; their MVN formulation accounts for correlation across these parameters in a random-effects fashion. Here, however, there is only one triple of sensitivity, specificity, and prevalence at each threshold, so there is no correlation estimand to be modelled. In fact, jointly modelling positive predictions and disease outcomes makes the parameters orthogonal by construction, and, hence, their estimators are also independent (see section \textit{\nameref{sec:bdca-details}}). On the other hand, the model from Wynants et al. (2018)\cite{Wynants_2018} benefits from partial pooling, being more appropriate for evidence synthesis in the multiple-setting context (in which case model (\ref{bayes-dca-model}) represents a complete pooling alternative). Finally, unlike sampling from (\ref{bayes-dca-model}), Wynants et al. (2018)\cite{Wynants_2018} method requires MCMC and is, therefore, expected to be considerably slower.

\begin{figure}[H]
\captionsetup{width=.95\linewidth}
\begin{center}
\includegraphics[width=.95\linewidth]{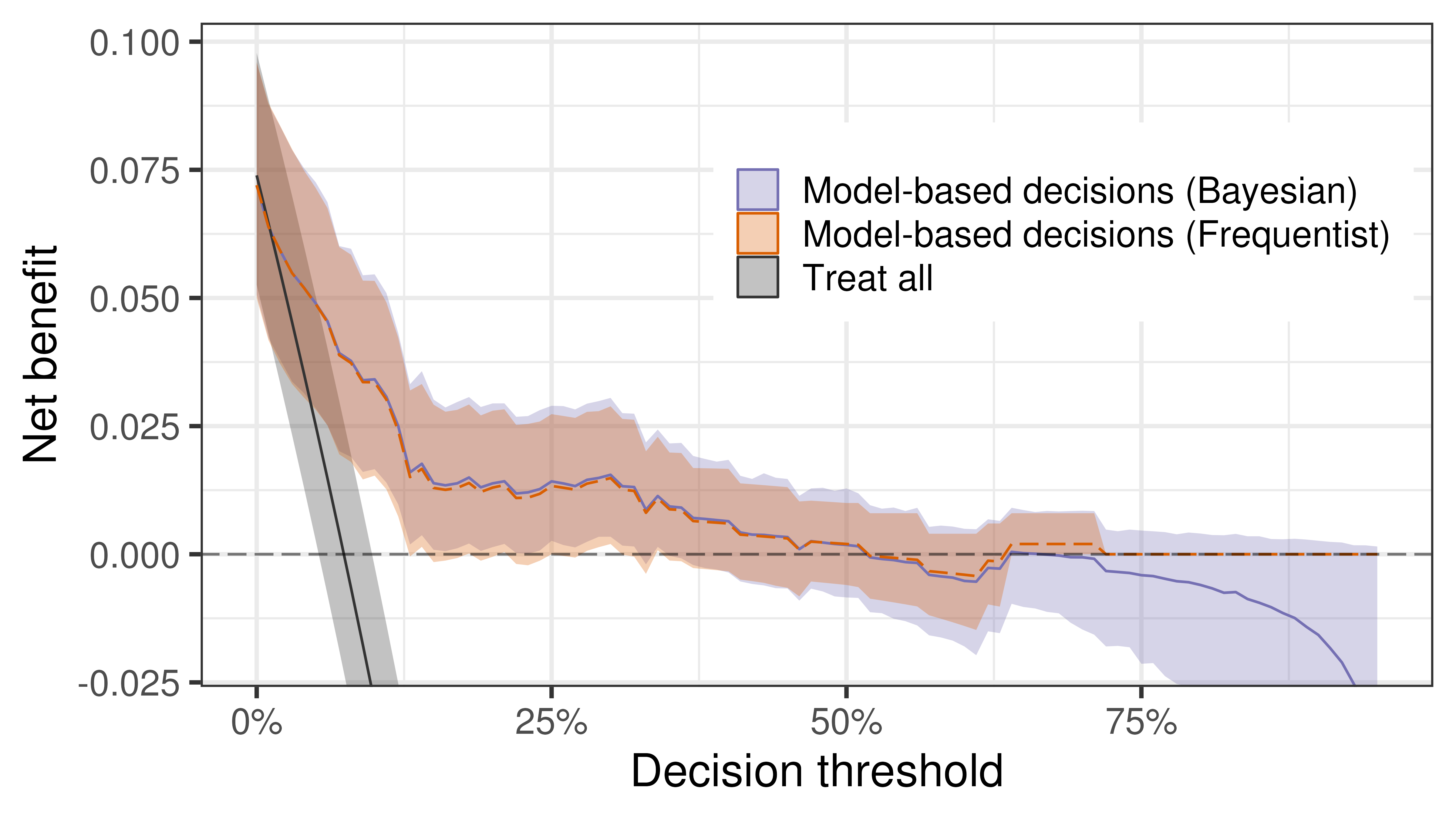}
\end{center}
\caption{\textbf{Bayesian DCA captures uncertainty across the entire decision curve.} Bayesian DCA was computed using the bayesDCA R package, while the Frequentist alternative used the bootstrap-based rmda package. An example model was built using data from the GUSTO-I trial\cite{gusto1993}, while DCA was constructed with held-out data ($N=500$, 36 events). The bootstrap intervals and point estimates collapse to zero as the threshold approaches the maximum risk prediction. The Treat all curve is nearly identical for both approaches and the Bayesian version is shown. Intervals correspond to 95\% confidence and credible intervals for Frequentist and Bayesian methods, respectively.}
\label{fig:gusto-example}
\end{figure}

We use data from the GUSTO-I trial\cite{gusto1993}, a large randomized study involving thrombolytic treatments for Acute Myocardial Infarction, as an example. Figure \ref{fig:gusto-example} shows the resulting Bayesian DCA with the Frequentist counterpart superimposed. Although point estimates mostly coincide, notice that the bootstrap-based DCA collapses to zero as the threshold increases -- in particular, as the threshold approaches the maximum observed risk prediction. The same does not happen with the Bayesian approach, which continues to naturally propagate uncertainty through the net benefit equation even in the absence of events, though with an increasing influence of the prior distribution. This behaviour can be useful, e.g., in settings with small effective sample sizes such as when most risk predictions are concentrated on one side of the decision threshold. While a decision threshold higher than all risk predictions in the population does imply zero net benefit, in small samples this may happen by chance alone -- in which case a positive net benefit would require more informative priors. Nonetheless, the bayesDCA R package warns the user if no events were observed above some decision threshold. Notice that, at very high thresholds, high sensitivity and specificity are required to yield a positive net benefit because of the increasing weight of false positives (i.e., large $w_t$). Thus, unless under strong prior knowledge or in face of substantial evidence of high sensitivity and specificity, the posterior distribution will tend toward net clinical harm in this region. The Bayesian DCA, therefore, indicates that harm is likely at higher thresholds for this example using vague priors.

Beyond uncertainty propagation, three key advantages come naturally with the Bayesian approach for binary outcomes implemented in bayesDCA. First, one can easily use full external information to estimate the prevalence parameter. In case-control studies\footnote{Notice, however, that non-nested case-control studies are at high risk of spectrum bias, potentially leading to an overestimation of sensitivity and specificity\cite{Usher_Smith_2016}.}, DCA needs to be adjusted for the population prevalence, which is usually done by plugging in a point estimate and completely ignoring uncertainty\cite{Marsh_2019}. Here, we may simply sample $p$ from the posterior (\ref{bayes-dca-model}) using data from an external cross-sectional study or the source population cohort. Equivalently, this could be seen as constructing an informative prior distribution from external prevalence information and sampling from the prior. Regardless of the source, the prevalence posterior distribution is then used to compute the net benefit. This takes into account the uncertainty in the prevalence parameter because we are using the raw prevalence data instead of plugging in a point estimate. The second advantage that our Bayesian approach brings is the option to use informative priors to improve estimation -- in terms of both uncertainty and point estimates. For instance, we know that $\textrm{Se}$ is higher for small thresholds and lower for large thresholds -- and that the reverse is true for $\textrm{Sp}$. The informative prior suggested in the Supplement takes advantage of this reasoning (section \textit{\nameref{supp:inf-priors}} and Supplementary Figures (\ref{fig:ppc}--\ref{fig:ppc2})). Additionally, prior elicitation is straightforward for binary tests since their sensitivity and specificity are fixed across thresholds. 

The third and most notable advantage of our Bayesian approach for DCA is the ability to interrogate posterior decision curves with an intuitive probabilistic interpretation. Since we have access to the full posterior distribution of the net benefit across all thresholds of interest, we can compute arbitrary functions to help us interpret the DCA output. Since this advantage applies to Bayesian DCA in general and not just to binary outcomes, we first propose a method of Bayesian DCA for survival outcomes and then describe the proposed probabilistic interpretation framework based on the interrogation of the posterior decision curves.

\subsection{Bayesian DCA for survival outcomes}
Many decision strategies address prognostic problems. For such survival outcomes, we must rewrite the net benefit formula to be able to account for censoring as follows
\begin{align}
    \textrm{NB}^\tau_t = \Big[1 - S\left(\tau \,\vert\,\hat{r}_{\tau} > t\right)\Big] \cdot \P\left[\hat{r}_{\tau}>t\right]
    - S\left(\tau \,\vert\,\hat{r}_{\tau} > t\right) \cdot \P\left[\hat{r}_{\tau}>t\right]
    \cdot w_t \label{nb-survival}
\end{align}
where $\hat{r}_{\tau}$ is the predicted risk of the event of interest at time $\tau$ (a probability in the case of a prognostic model and 0 or 1 in the case of a prognostic test). At time $\tau$ and threshold $t$, the probability of a positive prediction is given by $\P\left[\hat{r}_{\tau}>t\right]$, and $S\left(\tau \,\vert\,\hat{r}_{\tau} > t\right)$ is the survival probability given a positive prediction.

To estimate (\ref{nb-survival}), we jointly model survival times $T$ (with censoring indicators $C$) and the indicator of positive predictions $Z$ with Weibull and Bernoulli likelihoods, respectively (see section \textit{\nameref{sec:bdca-details}} for full model specification). Under independent priors, the posterior distribution factorizes due to parameter orthogonality as:
\begin{align}
\pi(p, \bm{\theta}_1 \vert \textrm{Data}) \propto 
\pi(p \vert \mathcal{D}_0) 
\times
\pi(\bm{\theta}_1 \vert \mathcal{D}_+)
\end{align}
where $\mathcal{D}_0 = \big\{Z_i\big\}_{i=1}^n$ is the set of positive prediction indicators, 
 $\mathcal{D}_+ = \big\{T_i, C_i\big\}_{i\in [n]: z_i=1}$ is the survival dataset for patients with positive predictions. 
 Here, $p = \P\left[\hat{r}_{\tau}>t\right]$ while $\bm{\theta}_1 = \big(\alpha_1, \sigma_1 \big)$ represents Weibull shape and scale parameters, respectively. Although the resulting posterior distribution does not have a closed form, it does benefit from orthogonality between the Weibull and the Bernoulli components, allowing us to estimate them separately. Hence, we put a Beta prior on $\theta_0$ (uniform by default, as before) to take advantage of conjugacy, while $\bm{\theta}_1$ is estimated with MCMC using Stan\cite{Carpenter2017}. In bayesDCA, the default priors for the Weibull parameters are \begin{align}
     &\alpha_1 \sim \textrm{Half-Student-t}(5, 0, 1.5)
     &\sigma_1 \sim \textrm{Half-Student-t}(30, 0, 100)
     \label{surv-priors}
 \end{align}
 These priors put a nearly equal prior probability on increasing and decreasing hazards and are largely vague with respect to scale. bayesDCA also allows user specification of the prior parameters above and provides a Gamma prior option instead of Half-Student-t. Once sampling is done, we then combine the draws from the posterior distributions of $\bm{\theta}_1$ and $\theta_0$ to compute the net benefit given in (\ref{nb-survival}).

We can now interrogate the posterior decision curves for all decision strategies under investigation. The entire interpretation framework proposed in the next section is immediately available for both survival and binary outcomes, highlighting once again the advantages of the proposed Bayesian approach.

\subsection{Probabilistic interpretation framework for Bayesian DCA}

The main advantage of Bayesian DCA is the ability to arbitrarily interrogate posterior distributions of decision curves. This allows for a probabilistic interpretation framework that helps understand the degree of uncertainty imposed by the currently available data on the observed decision curves. We may ask, for instance: what is the posterior probability that the model under investigation is useful at a given threshold? Following the definition in Wynants et al. (2018)\cite{Wynants_2018}, that is:
\begin{align}
    \textrm{P}\Big(\textrm{useful}\Big) = \textrm{P}\Big(NB_{\textrm{model}} > \max\big\{NB_{\textrm{treat all}}, NB_{\textrm{treat none}}\big\}\Big)\label{puseful}
\end{align}
where $NB_{\textrm{treat none}}$ is always zero. However, if we have two or more competing models, another natural question may be: what is the probability that my model is the \textit{best} decision strategy available? For instance, for a given ``model 1":
\begin{align}
\textrm{P}\Big(\textrm{best}\Big) = \textrm{P}\Big(NB_{\textrm{model}_1} > \max\big\{ NB_{\textrm{treat all}}, NB_{\textrm{treat none}}, NB_{\textrm{model}_2},
    NB_{\textrm{model}_3},\cdots \big\}\Big)\label{pbest}
\end{align}
Notice that Sadatsafavi et al. (2022)\cite{Sadatsafavi2023} define P(useful) as equation (\ref{pbest}), whereas here we define P(useful) and P(best) separately. This is because the best decision strategy might not be available everywhere, so the usefulness of remainder strategies is still relevant in that case. The above definitions can be extended to an arbitrary number of models or tests and be computed across all decision thresholds. Within bayesDCA, one may also compute pairwise comparisons between, say, two models with very similar net benefits. For instance, compute the probability that a given strategy beats another by at least $c$ net benefit units (i.e., net true positives):
\begin{align}
    \textrm{P}\Big(NB_{\textrm{model}_1} - NB_{\textrm{model}_2} > c \Big)
    \ \ \ \ \ \ \ \ 
    \textrm{ $c \geq 0$}
\end{align}
Upon full posterior interrogation, we may reach a better understanding of the net benefit profiles of the decision strategies under investigation: if there is large uncertainty around the decision curves, the above probabilities will be inconclusive (i.e., far below 100\%), so we may opt to collect more data before making a decision. 

One way to directly quantify the expected consequences of the current level of uncertainty is to compute the Expected Value of Perfect Information (EVPI) for model validation\cite{Sadatsafavi2023}:
\begin{gather}
EVPI = E_{\max} - \textrm{max}_{E}\\
\begin{aligned}
E_{\max} &= \mathbbm{E}\bigg[\max\Big\{ NB_{\textrm{treat all}}, NB_{\textrm{treat none}}, NB_{\textrm{model}_1}, NB_{\textrm{model}_2}, \cdots  \Big\}\bigg]\nonumber\\
    \textrm{max}_{E} &= 
    \max\bigg\{ 
    \mathbbm{E}\big[ NB_{\textrm{treat all}} \big],
    \mathbbm{E}\big[ NB_{\textrm{treat none}} \big],
    \mathbbm{E}\big[ NB_{\textrm{model}_1} \big],
    \mathbbm{E}\big[ NB_{\textrm{model}_2} \big],
    \cdots
    \bigg\}
\end{aligned}
\end{gather}
where the maximum in $E_{\max}$ is computed for each draw of the joint posterior distribution, whereas $\textrm{max}_E$ is a maximum of posterior means. The EVPI may be seen as the expected net benefit loss due to the current level of uncertainty in the decision curves. For instance, if the EVPI is 0.1, picking the \textit{observed} best decision strategy is associated with an expected loss of 0.1 net true positives as compared to picking the \textit{actual} best decision strategy.

It is important to notice another advantage of the parametric Bayesian approach to computing
the EVPI. As a sample statistic, the EVPI is expected to decrease monotonically with the sample
size\cite{Sadatsafavi2023}. In a simulation study using the GUSTO-I trial data, however, Sadatsafavi et al. (2023)\cite{Sadatsafavi2023}
showed that small effective sample sizes may cause the observed EVPI to escape this monotonic
behaviour, especially at very low or very high decision thresholds. One way to avoid this issue
is to employ informative prior distributions in Bayesian DCA. Reproducing the simulation code
from Sadatsafavi et al. (2023)\cite{Sadatsafavi2023}, we show how our Bayesian approach can recover the expected monotonic behaviour of the EVPI in the Supplement (section \textit{\nameref{supp:evpi-monotonic}} and Supplementary Figure (\ref{fig:evpi-simulation})).

In the next section, we provide a simulation study of the empirical performance of the suggested Bayesian approaches. In the following section, we present a case study to highlight the bayesDCA workflow, where we fully interrogate the posterior decision curves to quantify uncertainty around the answers to the four fundamental questions mentioned above. Here we focus on binary outcomes, though the same workflow for survival outcomes is also implemented and easily accessible through the bayesDCA R package.

\subsection{Empirical performance of Bayesian DCA estimation}

\subsubsection{Simulation study for binary outcomes}

To test the approach for binary outcomes, we simulate a population with an underlying logistic regression model and select an example model to be evaluated using DCA. To represent a range of scenarios, we vary the outcome prevalence and the maximum achievable discrimination, measured by AUC, representing the setting's signal-to-noise ratio. To resemble a common scenario of overfitting, the example model is miscalibrated: its predictions are overly extreme (i.e., too close to zero or to one). Each simulation run emulates a different external validation or test dataset selected at random from the setting's population, with which we perform DCA. The sample size for each simulated dataset is set so that the expected number of events is 100. See section \textit{\nameref{sec:sim-details}} for a full description of the simulations.

\begin{figure}[H]
\captionsetup{width=.95\linewidth}
\begin{center}
\includegraphics[width=.95\linewidth]{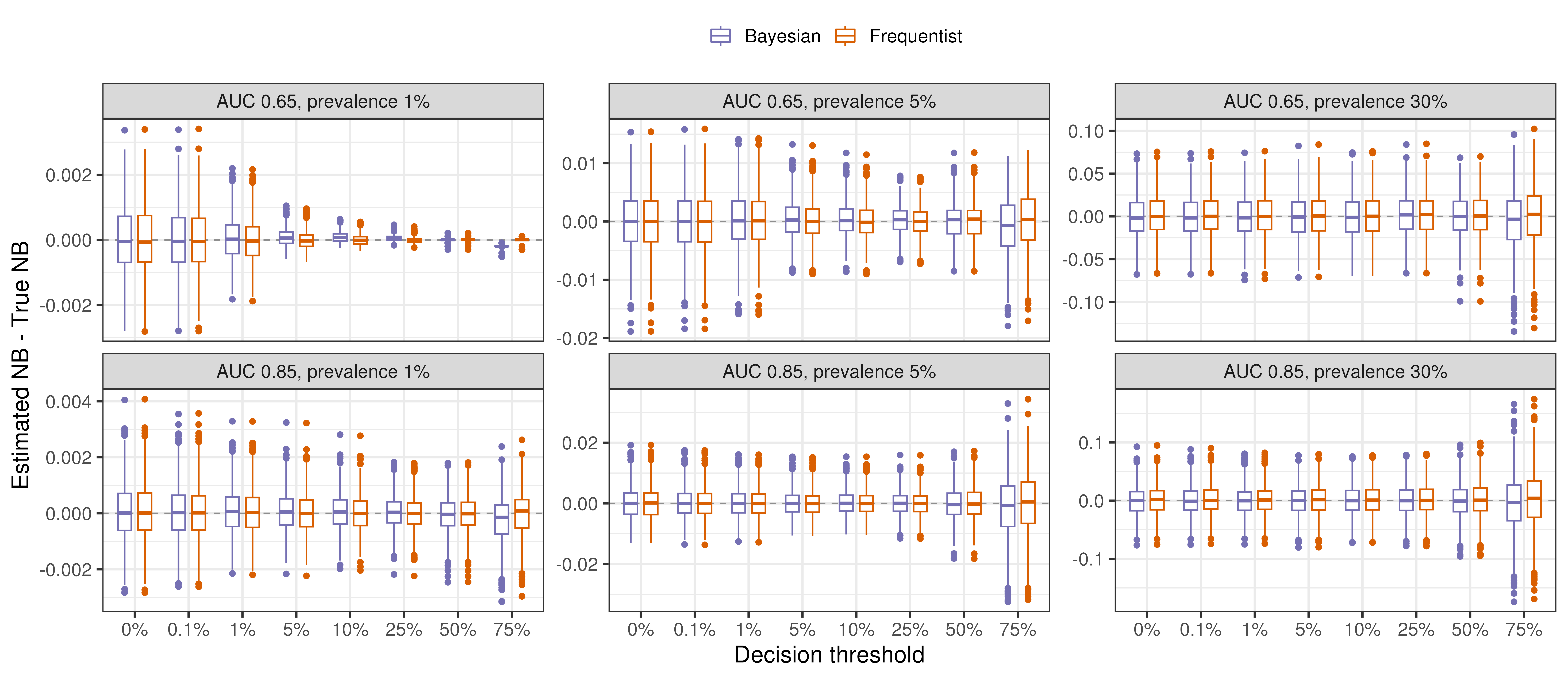}
\end{center}
\caption{\textbf{Bayesian and Frequentist DCA for binary outcomes show similar distributions of point estimate errors.} Bayesian DCA was computed using the bayesDCA R package, while the Frequentist alternative used the bootstrap-based rmda package. For each simulation run, DCA was performed for a fixed example model using a simulated test dataset of sample size corresponding to 100 expected events. A total of \simruns Monte Carlo repetitions was run for each setting. The setting AUC corresponds to its maximum achievable AUC. The example model for each setting was fixed to approximate the maximum discrimination of that setting but was miscalibrated (overly extreme risk predictions).}
\label{fig:point-estimates-error}
\end{figure}

For each setting, we ran \simruns simulations performing both Bayesian DCA and bootstrap-based Frequentist DCA for comparison. Figure (\ref{fig:point-estimates-error}) shows the resulting distributions of estimation errors. As expected, the distributions of point estimate errors are nearly identical for the Bayesian and the Frequentist approaches in almost all simulation settings and decision thresholds. In general, with 100 expected events, any influence of the default vague priors is unnoticeable. One exception is the setting with an AUC of 0.65 and a prevalence of 1\%, in particular at a threshold of 75\%. Given the low prevalence and discrimination, very high thresholds such as 75\% often yield no positive prediction (i.e., no risk prediction above the decision threshold), so the effective sample size for that threshold is low and the prior matters more. This is the setting in which bootstrap-based point estimates and intervals are expected to collapse to zero, while the Bayesian estimates are regularized by the prior distribution. Still, the discrepancy between the two approaches is negligible in absolute terms -- see Supplementary Figure (\ref{fig:point-estimates-distribution}) for a plot of the point estimates in the original scale with the true net benefit overlayed.

\begin{figure}[H]
\captionsetup{width=.95\linewidth}
\begin{center}
\includegraphics[width=.95\linewidth]{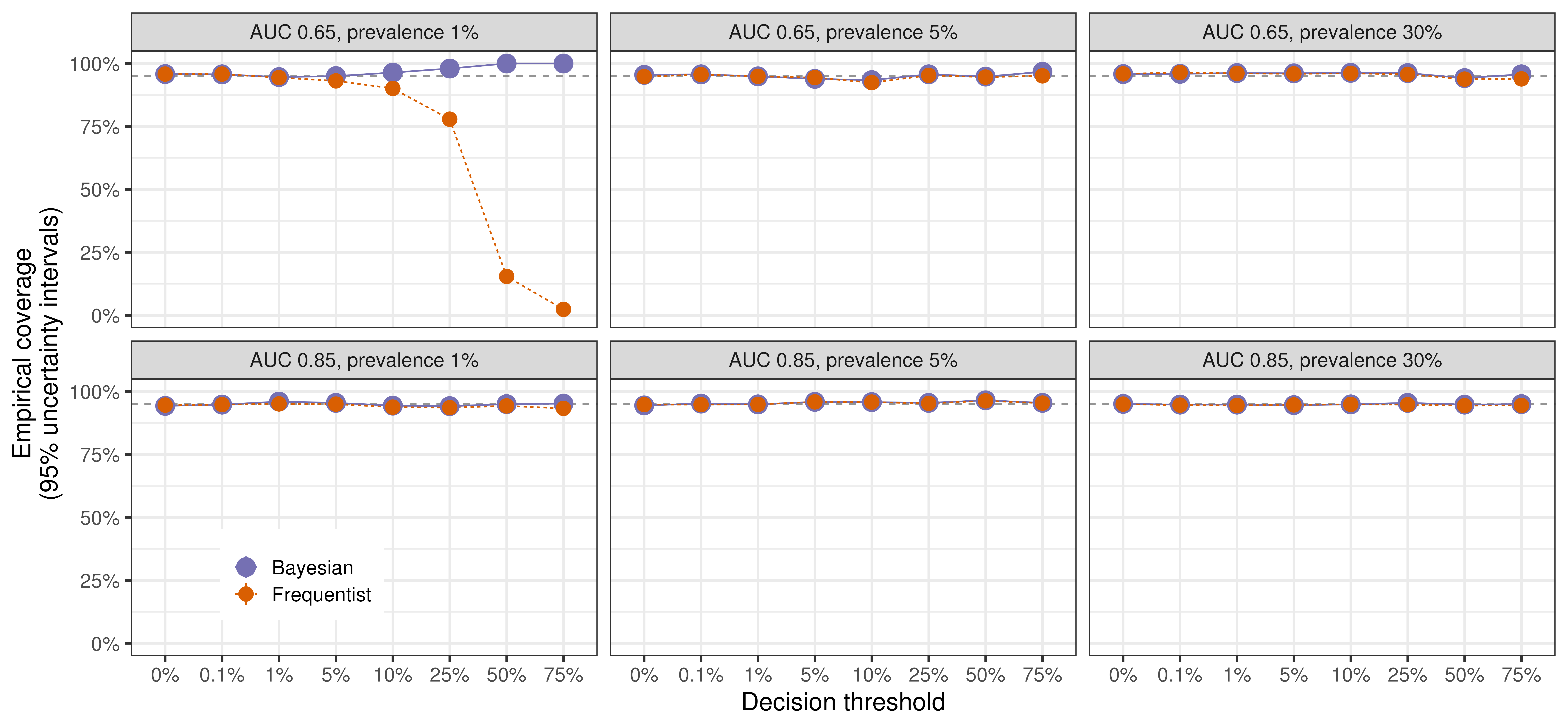}
\end{center}
\caption{\textbf{Bayesian and Frequentist DCA for binary outcomes show calibrated uncertainty intervals.} For both approaches, coverage matches nominal values in almost all cases. The points of miscalibration (AUC 0.65, prevalence 1\%, thresholds of 10\% or above) are due to very low rates of positive predictions (i.e., above the corresponding threshold). In this scenario, bootstrap intervals collapse to zero so extreme undercoverage is observed; Bayesian intervals are more dependent on the prior distribution, showing minimal overcoverage. Bayesian DCA was computed using the bayesDCA R package, while the Frequentist alternative used the bootstrap-based rmda package. For each simulation run, DCA was performed for a fixed example model using a simulated test dataset of sample size corresponding to 100 expected events. A total of \simruns Monte Carlo repetitions was run for each setting. The setting AUC corresponds to its maximum achievable AUC. The example model for each setting was fixed to approximate the maximum discrimination of that setting but was miscalibrated (overly extreme risk predictions).}
\label{fig:emprirical-coverage}
\end{figure}

We also assessed the empirical coverage of 95\% uncertainty intervals for both Bayesian and Frequentist methods. Although not required for a valid Bayesian analysis, users of the bayesDCA R package may feel more comfortable having Frequentist calibration of credible intervals (Cr.I.), at least under vague priors. As shown in Figure (\ref{fig:emprirical-coverage}), the empirical coverage of both the Bayesian and Frequentist decision curves closely matches the nominal value of 95\% for the entire range of decision thresholds in most simulation settings. The only exception is, once again, the setting with an AUC of 0.65 and a prevalence of 1\%. At thresholds of 10\% or above, the bootstrap intervals show increasing undercoverage, reaching an empirical coverage of nearly 0\% for the 75\% threshold. This happens because although the true net benefit is not exactly zero in this setting due to the presence of positive predictions in the population, the bootstrap intervals tend to collapse to zero due to the lack of positive predictions in the observed samples. In contrast, under the same problematic scenario, Bayesian intervals show only a slight overcoverage. Despite its vagueness, the default prior distribution yields valid credible intervals with reasonable Frequentist calibration. Moreover, the Bayesian credible intervals are not significantly wider than the Frequentist confidence intervals, except when the Frequentist method fails. Overall, Supplementary Figure (\ref{fig:ci-width}) shows that the Bayesian intervals have reasonable width, even when the Frequentist counterpart collapses to zero.

Finally, as shown in Supplementary Figure (\ref{fig:runtime}), the Bayesian DCA for binary outcomes is orders of magnitude faster than its bootstrap-based Frequentist counterpart. Here, we are sampling 4000 draws from the posterior distribution (\ref{bayes-dca-model}) and using 500 bootstrap samples -- the default in bayesDCA and rmda, respectively. Moreover, computation time significantly increases with the overall sample size for the bootstrap case, but not in the Bayesian case. Given the 100 expected events fixed for each simulation run, simulation settings with prevalences of 30\%, 5\%, and 1\% imply overall sample sizes around 333, 2,000, and 10,000, respectively. While this is expected to impact bootstrap speed, the computation time for the proposed Bayesian DCA is virtually unaffected because (\ref{bayes-dca-model}) only depends on simple summary statistics. In our experience, running Bayesian DCA for binary outcomes with bayesDCA takes no more than a second in most cases (using a standard laptop with 12GB of RAM and no parallelization).

\subsubsection{Simulation study for survival outcomes}
We follow the same simulation strategy as above. The underlying populations follow Weibull distributions with covariates satisfying the proportional hazards (PH) assumption, while censoring times are uniformly distributed. The decision strategies being investigated are fixed PH Cox models with exaggerated coefficients -- i.e., miscalibrated due to overfitting. For each setting, we simulate \simruns datasets, with which we perform both Bayesian and Frequentist DCA using a twelve-month prediction horizon. We vary the underlying C-statistics and one-year survival rate. For brevity, we show the results for C-statistics 0.6 and 0.9 with a one-year survival of 10\%. A full description of the simulations is provided in the section \textit{\nameref{sec:sim-details}}, and further simulation settings are reported in the Supplementary Figures (\ref{fig:point-estimates-error-surv}--\ref{fig:surv-mape}).
\begin{figure}[H]
\captionsetup{width=.95\linewidth}
\begin{center}
\includegraphics[width=.95\linewidth]{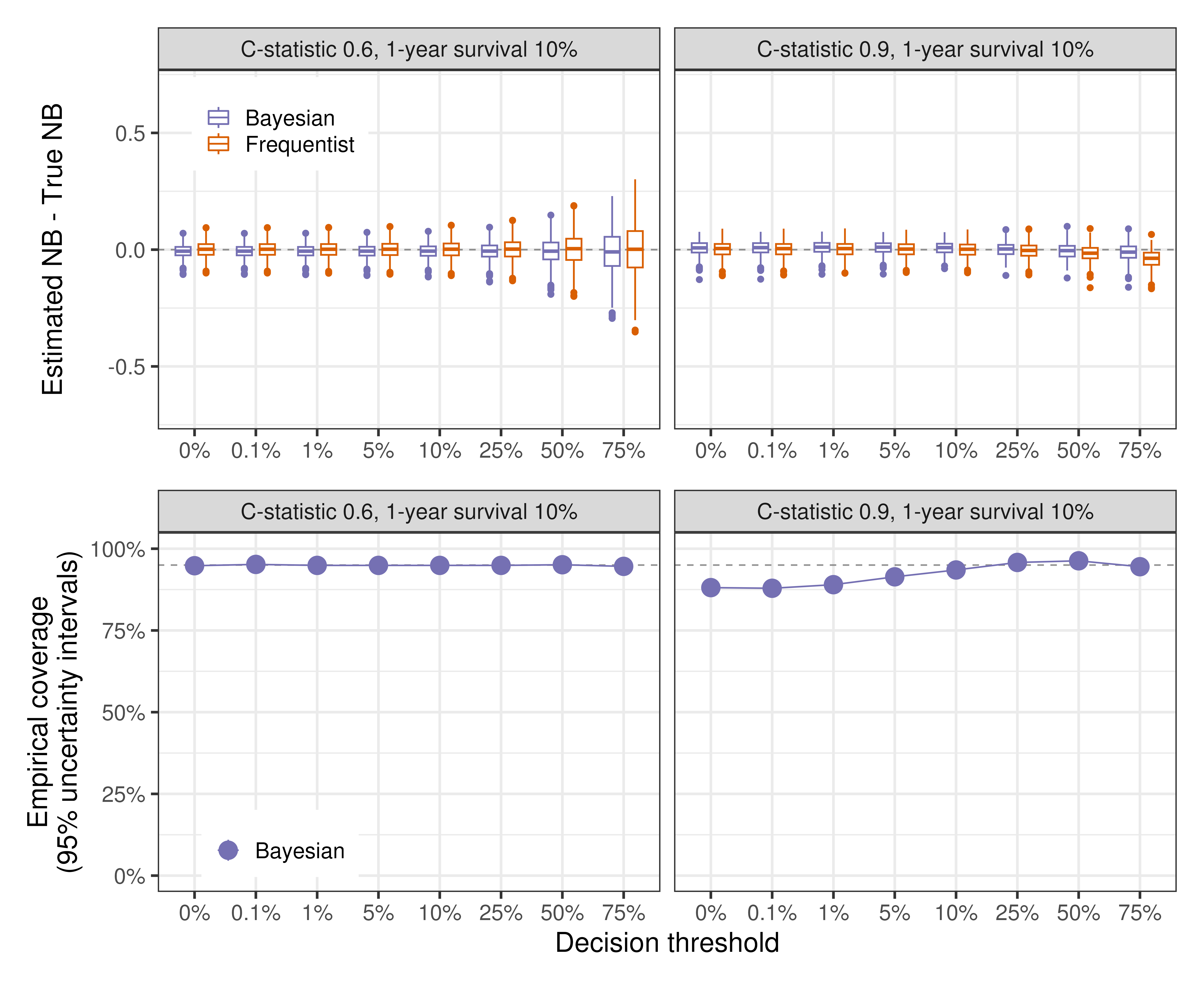}
\end{center}
\caption{\textbf{Bayesian and Frequentist DCA for survival outcomes show similar estimation performance.} Bayesian DCA was computed using the bayesDCA R
package, while the Frequentist alternative used the dcurves package. For each simulation run, DCA was performed for a fixed example model using a simulated test dataset of sample size corresponding to 100 expected events. A total of \simruns Monte Carlo repetitions was run for each setting. The setting C-statistic corresponds to its maximum achievable discrimination. The example model for each setting was fixed to approximate the maximum discrimination of that setting but was miscalibrated (overly extreme risk predictions). We did not find an easily-accessible Frequentist implementation of survival DCA with uncertainty intervals and, therefore, empirical coverage is shown for the Bayesian approach only. The prediction horizon is one year. Further simulation settings are in the Supplementary Figures (\ref{fig:point-estimates-error-surv}--\ref{fig:surv-mape}).}
\label{fig:survival-sim}
\end{figure}

As shown in Figure (\ref{fig:survival-sim}), the Bayesian point estimates generally behave similarly to the ones from the Frequentist approach. At decision thresholds of 50\% or above, the effective sample size can be small which causes a slight bias for both approaches depending on the simulation setting -- here seen at the 75\% threshold for the Frequentist approach under C-statistic 0.9, see also Supplementary Figure (\ref{fig:point-estimates-error-surv}). Because we are not aware of any Frequentist implementation of DCA for survival outcomes that provides uncertainty intervals, we show here coverage results for the Bayesian approach only. The Bayesian uncertainty intervals show reasonable empirical coverage overall, though we observed undercoverage for very high or very low thresholds depending on the simulation setting. While the setting with C-statistic 0.9 shown in Figure (\ref{fig:survival-sim}) represents the worst-case scenario, most coverage probabilities remained above 90\% across all settings -- Supplementary Figure (\ref{fig:empirical-coverage-surv}). Mean absolute percentage errors of point estimates were comparable between Bayesian and Frequentist approaches across all simulation settings -- Supplementary Figure (\ref{fig:surv-mape}).

In summary, our Bayesian approach offers an accurate alternative for estimating decision curves for both binary and survival outcomes. It also enables uncertainty quantification for survival outcomes, and provides much faster quantification of uncertainty for binary outcomes. These methods are implemented and easily accessible in the bayesDCA R package. While the simulations shown in this section apply the default weakly-informative priors, different priors may be specified by the user if desired to further improve estimation performance. The bayesDCA R package allows sampling from the prior only so that prior predictive checks are straightforward.

\subsection{Applied case study}
Clinical prediction models are commonly employed to predict cancer diagnosis. For example, the ADNEX model predicts an individual's risk of having ovarian cancer with high discrimination (AUC>0.9) and adequate calibration \cite{Van_Calster_2014}. The model employs clinical and ultrasound features from patients under suspicion of ovarian cancer due to the known or suspected presence of adnexal masses. Previously, 
Wynants et al. (2019) warned about the importance of utility-based decision thresholds and used the ADNEX model as a motivating example \cite{Wynants_2019}. The authors suggest that a reasonable decision threshold may be as low as 6\% due to the high cost of false negatives which can cause late detection and treatment of aggressive cancer. Here, we will expand on their example using Bayesian DCA on a hypothetical scenario.

\begin{figure}[H]
\captionsetup{width=.95\linewidth}
\begin{center}
\includegraphics[width=.95\linewidth]{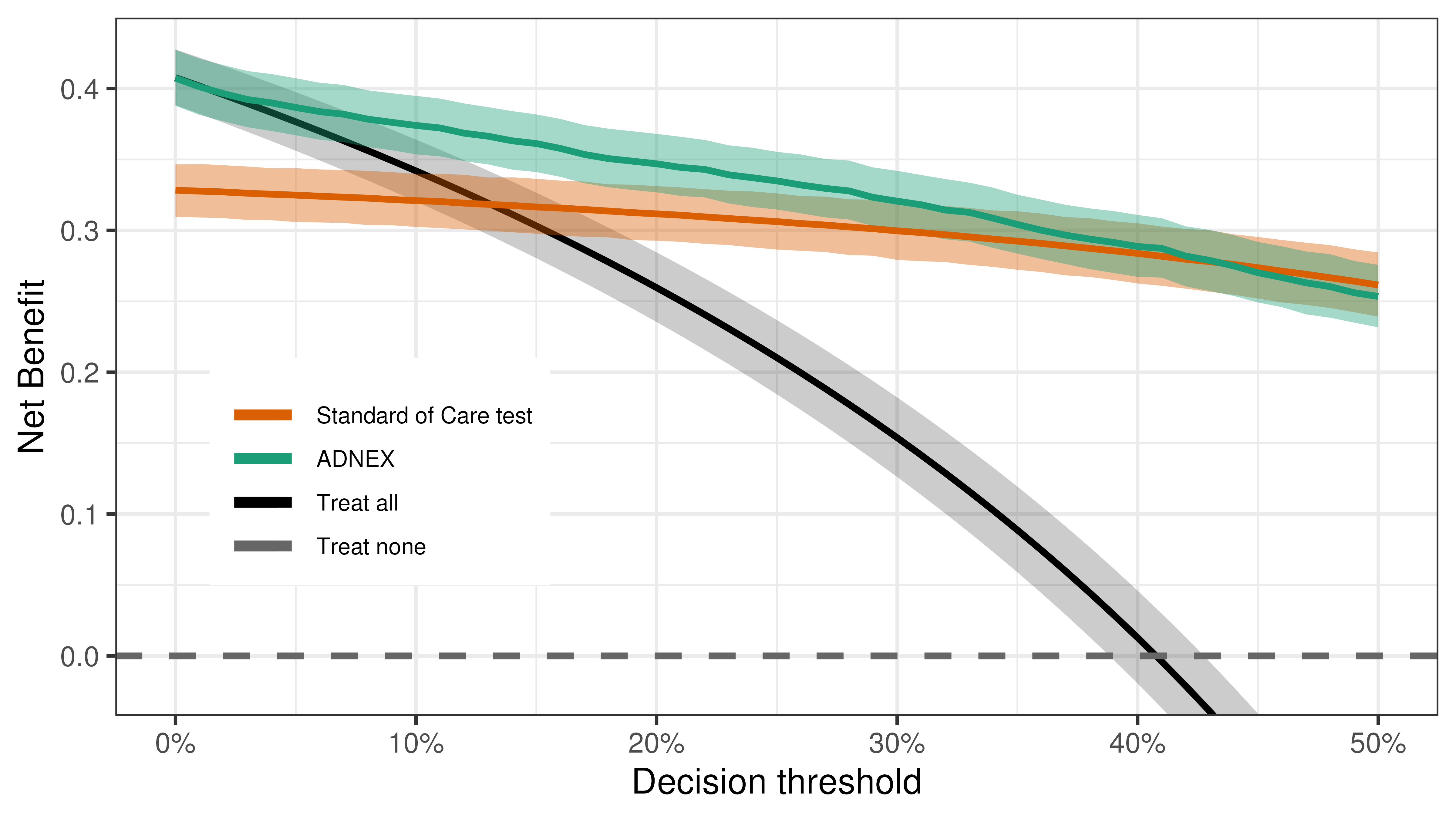}
\end{center}
\caption{\textbf{Illustration of Bayesian DCA for the ADNEX model and hypothetical Standard of Care diagnostic test.} Bayesian DCA was computed using the bayesDCA R package and publicly-available data from Wynants et al. (2019)\cite{Wynants_2019} ($N=2403$, 980 events). A hypothetical Standard of Care diagnostic test was simulated to have 81\% sensitivity and 88\% specificity. Lines are posterior means and uncertainty intervals are 2.5\% and 97.5\% posterior percentiles (i.e., 95\% credible intervals).}
\label{fig:dca-case-study}
\end{figure}

Suppose we perform an external validation study to assess the predictive performance of the ADNEX model. We wish to know whether the model should replace the Standard of Care (SoC) currently in place: a hypothetical diagnostic test with 81\% sensitivity and 88\% specificity. Using the publicly-available data from Wynants et al. (2019)\cite{Wynants_2019} ($N=2403$, 980 events), the Bayesian DCA results are shown in Figure (\ref{fig:dca-case-study}).

In terms of the point estimate of net benefit, the original ADNEX model is superior for most decision thresholds: under no additional costs to implement or use, ADNEX-based decisions would be the best strategy to follow. However, there is substantial uncertainty around the clinically-motivated threshold of 6\%, where it is not immediately clear if the ADNEX superiority is simply due to chance. For example, implementing the model into clinical practice and using it as part of daily care may impose challenges on healthcare institutions with limited resources. Moreover, we may require a low degree of uncertainty before replacing the current Standard of Care (or default strategies) to prevent having to undo the implementation of a suboptimal strategy. Thus, inspecting the DCA alone, under the threshold of 6\%, we might require more evidence confirming ADNEX superiority before implementing it into clinical practice. 

Additionally, uncertainty intervals from the ADNEX model and the hypothetical SoC test start overlapping at higher decision thresholds around 40\%. The SoC test becomes superior in terms of estimated net benefit at very high thresholds. Here, we consider only decision thresholds below 50\% due to the assumption that, in the case of ovarian cancer diagnosis, the cost of false negatives is generally larger than the cost of false positives.

\begin{figure}[H]
\captionsetup{width=.95\linewidth}
\begin{center}
\includegraphics[width=.95\linewidth]{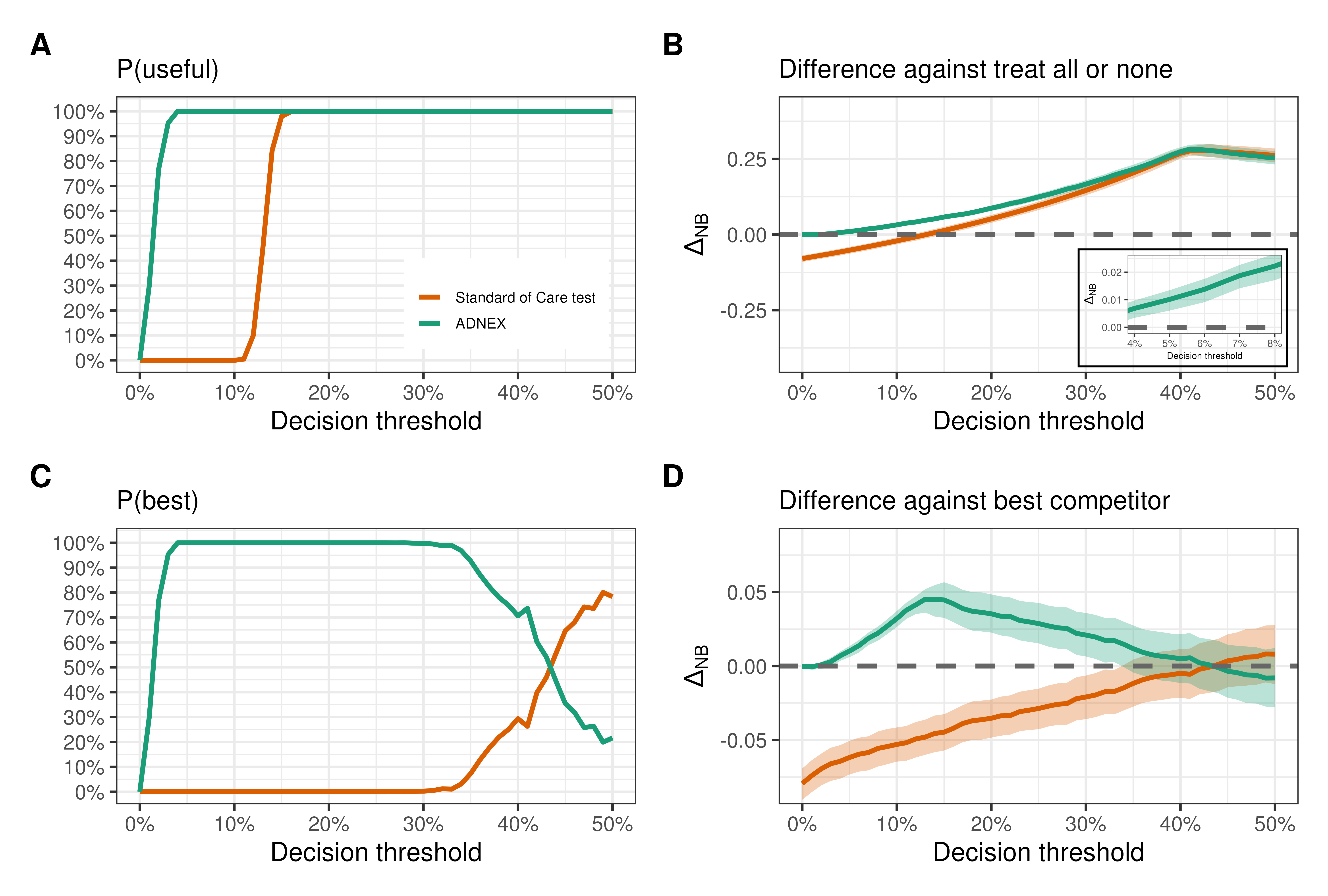}
\end{center}
\caption{\textbf{Bayesian DCA estimates the probability that each decision strategy is useful or the best among all strategies considered.} Bayesian DCA was computed using the bayesDCA R package ($N=2403$, 980 events). \textbf{(A)} P(useful) is the probability that a given decision strategy has a higher net benefit than the Treat all and Treat none strategies (i.e., beats the default strategies) and is computed from \textbf{(B)} the difference in NB between each strategy and Treat all/none. \textbf{(C)} P(best) is the probability that a given strategy has a higher net benefit than all the remaining strategies (i.e., beats its best competitor) and is computed from \textbf{(D)} the difference in NB between each strategy and the maximum net benefit among other competing strategies.}
\label{fig:dca-posterior-interrogation}
\end{figure}

From its decision curve alone, it is clear that the ADNEX model is clinically useful for higher thresholds, but there is considerable uncertainty at thresholds below 10\%. Besides visual inspection of the decision curves, how can we quantify our uncertainty about which decision strategies are clinically useful? We can answer this question by interrogating the posterior distribution: at the clinically-motivated threshold of 6\%, there is over 99.9\%  posterior probability that the ADNEX model is clinically useful -- Figure (\ref{fig:dca-posterior-interrogation}A). As the threshold increases, this posterior probability is consistently close to 100\%, and the SoC test becomes useful as well. Importantly, we are only able to speak of P(useful) at all and inspect posterior distributions for any decision strategy because we are adopting a Bayesian approach \cite{Sadatsafavi2023}. No natural counterpart exists under the Frequentist bootstrap. 
The results above may be counter-intuitive due to the overlapping uncertainty intervals in Figure (\ref{fig:dca-case-study}). However, notice that P(useful) at 6\% for the ADNEX model is determined by the difference between its net benefit and the one from the Treat all strategy -- Figure (\ref{fig:dca-posterior-interrogation}B), also notice the inset element within the plot for a zoomed view at the 6\% threshold. Their posterior distributions are highly correlated (R=0.98) due to their shared prevalence parameter, so the estimated net benefit difference is very precise: 0.014 (95\% Cr.I. 0.009 --- 0.018). However, to justify the implementation of the ADNEX model at the 6\% threshold, we must be confident that this gain in net benefit also overcomes any additional cost of using the model in daily practice, if this isn't already factored in. Since the strategy cost may depend on the local context, we simply report the estimated net benefit difference and its uncertainty. The consumer of the DCA results can then reason if they are confident enough that this estimated difference overcomes their context-specific costs. If not, more data may still be required prior to implementation.

Although usefulness is important (i.e., being better than Treat all and Treat none), ideally we would like to use the best decision strategy available. Among the strategies under consideration, we can once again interrogate the posterior distribution to quantify our uncertainty about what is the best decision strategy at each decision threshold -- Figures (\ref{fig:dca-posterior-interrogation}C) and (\ref{fig:dca-posterior-interrogation}D). We do this by comparing each strategy against the best of the remaining strategies (i.e., the ``best competitor"). For a given strategy at a given threshold, if the difference in net benefit against its best competitor is positive, then this strategy is better than all remaining strategies.

For instance, at the 6\% threshold, the best competitor against the ADNEX model is the Treat all strategy -- see Figure (\ref{fig:dca-case-study}). In this case, therefore, P(best) and P(useful) coincide: over 99.9\%. As expected, the probability that the ADNEX model is the best available strategy is virtually 100\% for most thresholds. For higher thresholds, however, there is increasingly more overlap between the uncertainty intervals from the ADNEX model and the SoC test. This translates into an increasingly higher P(best) for the SoC test and progressively lower P(best) for the ADNEX model. At the 50\% threshold, there is 84\% posterior probability that the SoC test is the best decision strategy available.

\begin{figure}[H]
\captionsetup{width=.95\linewidth}
\begin{center}
\includegraphics[width=.95\linewidth]{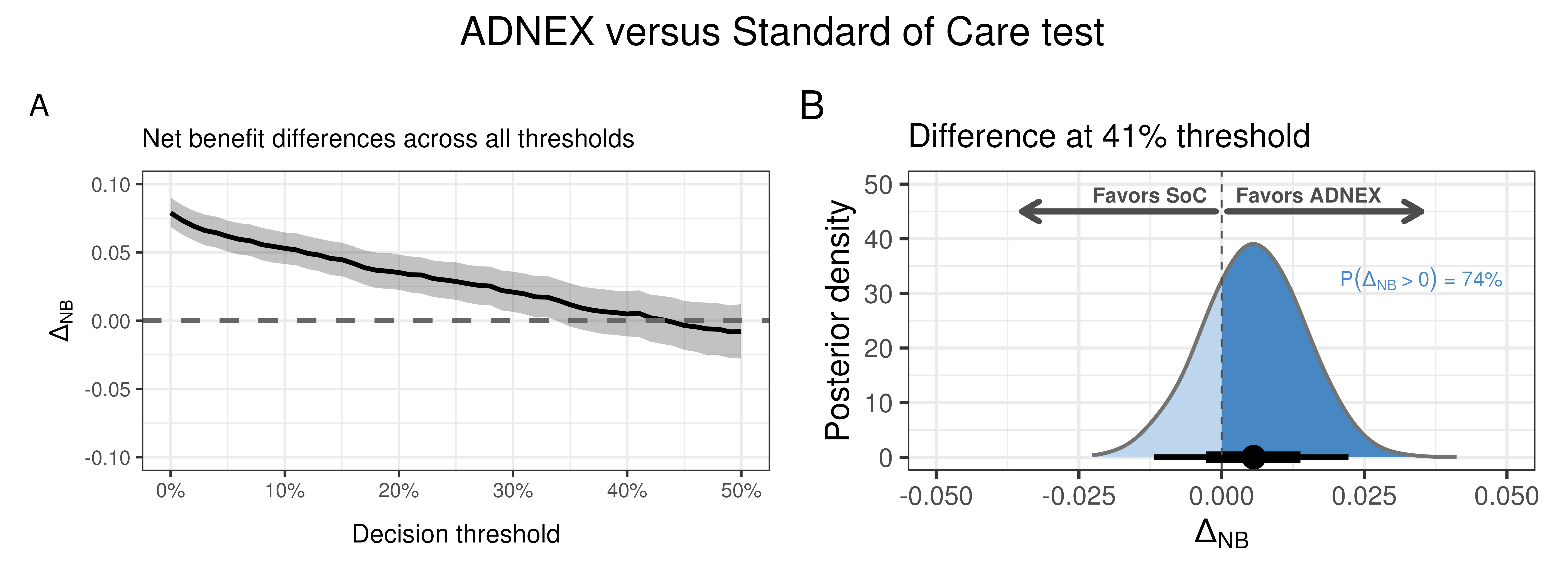}
\end{center}
\caption{\textbf{Pairwise net benefit comparison between the ADNEX model and hypothetical SoC test.} \textbf{(A)} Estimated difference in net benefit between ADNEX and the SoC test (positive y-axis favors ADNEX). \textbf{(B)} Posterior distribution of the difference for the specific threshold of 41\% (observed prevalence), at which there is a 74\% posterior probability that the ADNEX model is superior to the hypothetical SoC test.}
\label{fig:dca-pairwise}
\end{figure}

From Figures (\ref{fig:dca-case-study}) and (\ref{fig:dca-posterior-interrogation}), using the ADNEX model or treating all the patients is likely the best we can do under low decision thresholds. However, the decision-maker may be interested in higher thresholds. For instance, we may classify patients as high-risk if their predicted probability of cancer is higher than the prevalence in the present dataset (41\%)\cite{Wynants_2019}. The motivation for this is to maintain the proportion of high-risk patients, according to the model predictions, close to the actual disease prevalence\cite{Kelly_2008}. From Figure (\ref{fig:dca-posterior-interrogation}), we see that P(best) for the ADNEX model at the 41\% threshold is far less convincing: around 74\%. Moreover, the difference in net benefit between the ADNEX model and its best competitor (the SoC test in this case) is almost unnoticeable. Since we have full posterior distributions, we can directly compute a pairwise comparison between the ADNEX model and the SoC test, which is shown in Figure (\ref{fig:dca-pairwise}).
The superiority of the ADNEX model over the SoC test is clear for small thresholds. However, for high thresholds, the difference is smaller. At the 41\% threshold, the estimated difference is 0.003 (95\% Cr.I. -0.014 --- 0.019). At this threshold, the SoC test is the best competitor against the ADNEX model so the probability that the ADNEX model is better than the SoC test is again 74\% -- matches the corresponding P(best) for ADNEX. If no additional strategy costs are considered, maximizing observed net benefit would lead us to favor the implementation of the ADNEX model. However, there is still a 100\%--74\% = 26\% posterior probability that the current Standard of Care is better than the model for this decision threshold. In face of such large uncertainty, risk aversion may lead us to oppose model implementation unless more data is made available.

\begin{figure}[H]
\captionsetup{width=.95\linewidth}
\begin{center}
\includegraphics[width=.95\linewidth]{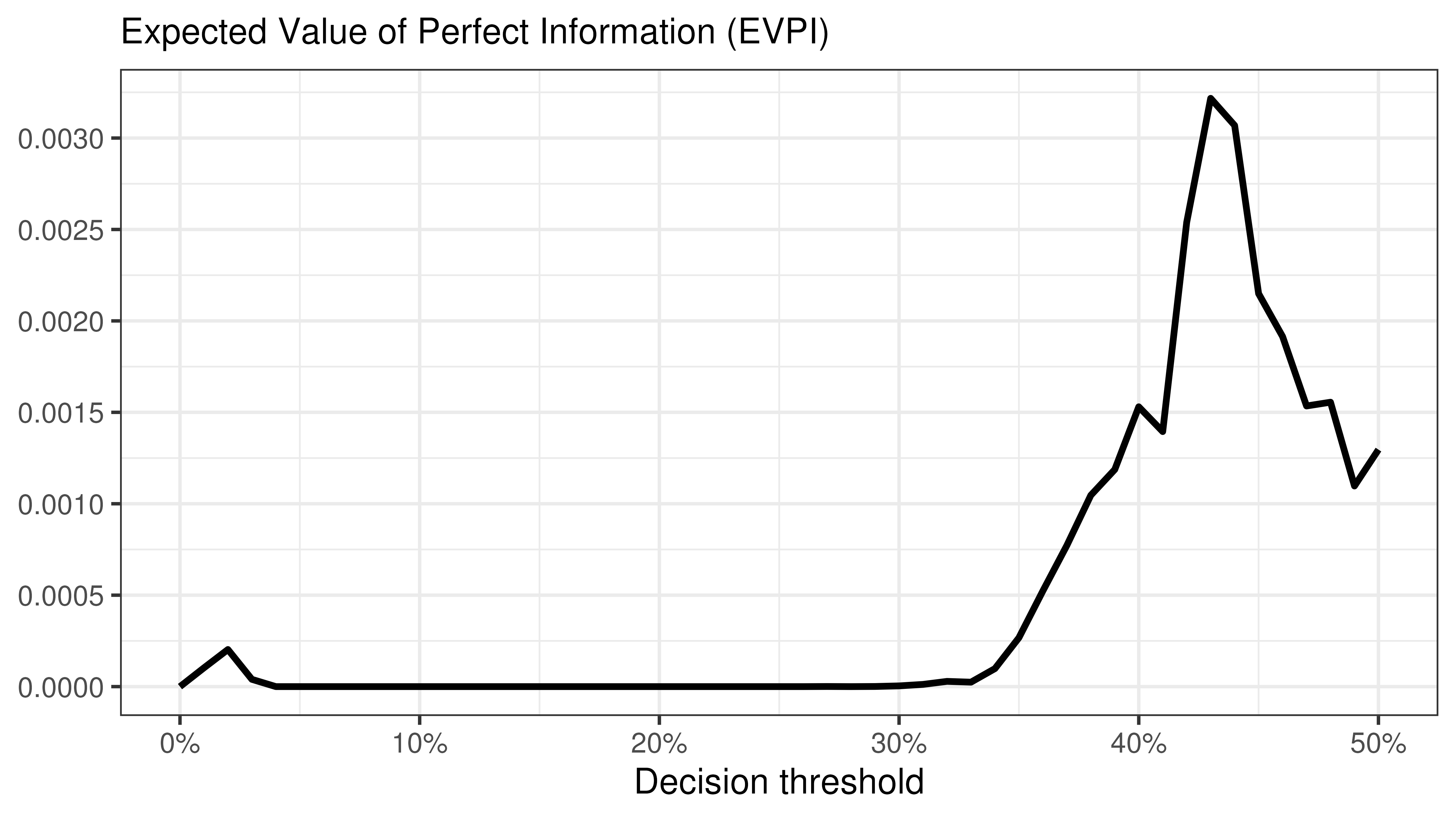}
\end{center}
\caption{\textbf{Consequence of the current level of uncertainty in the decision curves from the case study.} The validation Expected Value of Perfect Information (EVPI) shows the expected loss in net true positives due to estimation uncertainty.}
\label{fig:evpi}
\end{figure}

Finally, one might wonder if more data are needed to fully characterize the best decision strategy across all thresholds in the target population. It might be that the consequences of uncertainty at, e.g., very high thresholds are too costly in terms of net benefit. With more data, maybe we could confirm if the ADNEX model is indeed superior at the 41\% threshold, for instance. To directly quantify the consequences of the current level of uncertainty, Figure (\ref{fig:evpi}) shows the validation EVPI for the present case study. 
As a summary of the posterior distributions, the EVPI shows small, noisy variations. The highest EVPI across all the decision thresholds of interest is around 0.003. Whether this value is relevant depends, again, on the specific context\cite{Sadatsafavi2023}. For instance, if 1000 clinical decisions regarding the presence of ovarian cancer are made every year for a given population, then the current level of uncertainty is associated with missing at most three net true positive cases of ovarian cancer per year. If 10,000 decisions are made yearly, we could be missing out on up to 30 net true positives every year.

In summary, using the ADNEX model is the best decision strategy across small thresholds assuming no additional strategy costs according to the data. If there exists additional cost of using the model and we accept the decision threshold of 6\%, treating all patients or gathering additional data may be preferred. For the 41\% threshold, even though the net benefit point estimate for the ADNEX model is slightly higher than for the SoC test, the uncertainty may be too high to justify replacing the Standard of Care currently in place if we are not risk-neutral. For decision thresholds of 42\% and higher, we are increasingly confident that the current Standard of Care is the best decision strategy. Finally, collecting more data from the same population is expected to yield substantial net benefit gain if thousands of clinical decisions are made every year.

\section{Discussion}
Bayesian decision curve analysis was first proposed in the context of meta-analysis\cite{Wynants_2018}. An immediate advantage was the possibility of calculating P(useful), the probability that a decision strategy is clinically useful. The concept was later extended to the probability of being the best decision strategy available\cite{Sadatsafavi2023}. Here, we attempt to clarify terminology by defining the latter as P(best) instead -- which differs from P(useful) if we are evaluating more than one predictive model or test at the same time.

The key advantage of Bayesian DCA is the ability to fully interrogate posterior decision curves with an intuitive probabilistic interpretation. We may compute multiple quantities of interest to quantify uncertainty around the answers to fundamental questions such as which decision strategies are useful, what is the best decision strategy, make pairwise comparisons between strategies, and what is the expected consequence of the current level of uncertainty in terms of expected loss in net benefit. These may help the interpretation of DCA results: if the uncertainty is too large, then more data may be needed to properly choose strategies.

On the other hand, the bootstrap-based approaches for DCA currently implemented are limited to estimating confidence intervals around net benefit \cite{rmda2018}. Moreover, the coverage of these intervals is limited by the distribution of the predictions: under a low effective sample size, the bootstrap fails. This may pose a problem when the observed risk predictions are concentrated on one side of a given decision threshold (typically very high or very low thresholds), ultimately hiding the possibility of net harm. A low effective sample size may also cause the EVPI to misbehave, varying non-monotonically with the overall sample size -- see Figure 5 in Sadatsafavi et al. (2022)\cite{Sadatsafavi2023}. Bayesian DCA offers an alternative to potentially overcome these limitations.

From a Frequentist perspective, the main limitations of Bayesian DCA are the choice of priors and the interpretation of the uncertainty intervals. We chose weakly-informative priors as default in bayesDCA and provided a simulation study to address this valid concern. Under these priors, Bayesian intervals for the binary outcomes case showed near-perfect empirical coverage, with similar width to the bootstrap intervals. When stronger priors exist (e.g., from past studies), this knowledge may be used to obtain more accurate estimates. For the survival outcomes case, though empirical coverage of the Bayesian approach was reasonable overall, we did observe slight undercoverage in some simulation settings. This might be due to known biases in the estimation of the shape parameter of the Weibull distribution\cite{Makalic_2023}. Future research is warranted to further improve this method, potentially involving reparametrization of the Weibull likelihood to further leverage parameter orthogonality\cite{Yang_2003}. Informative prior elicitation for Bayesian DCA, in both binary and survival cases, may also be further developed in future work.

Another possible limitation of the Bayesian approach is the computation time due to sampling. In the binary outcomes case, we leveraged parameter orthogonality and posterior independence to allow particularly fast estimation. The bayesDCA implementation is multiple times faster than its bootstrap counterpart and typically takes less than a second, even for large sample sizes. In the survival outcomes case, we do need to use MCMC to estimate the Weibull parameters and, hence, the Bayesian approach is slower than in the binary outcomes case. From our experience, a typical DCA may take from three to five minutes in this case (using a standard laptop with 12GB of RAM and 8 cores, and running four MCMC chains in parallel taking 4000 draws each).

There is debate regarding the value of uncertainty quantification in DCA\cite{Vickers_2023}. From the risk-neutral point of view, it informs whether more data is warranted to confidently compare decision strategies, but wouldn't alter our choice of strategy given the data that is available at the moment of the clinical decision. However, when it comes to new decision strategies, the decision to replace the well-accepted Standard of Care may warrant a more conservative approach \cite{Vickers_2008, Glynn_2023}. Implementing a suboptimal strategy poses costs that are potentially irrecoverable. Challenges such as infrastructure development, physician adoption, and regulatory approvals cannot be frequently reversed. Importantly, patients harmed by new technologies due to premature implementation may face serious unwanted consequences. This is not to advocate for questionable inferential procedures such as threshold-specific p-values but to point out that caution may be warranted against the implementation of a new strategy that alters well-established clinical practice when, for instance, its P(useful) varies just above 50\% for a range of decision thresholds of interest -- the risk-averse approach.

Nonetheless, the present work proposes a Bayesian approach for the estimation of net benefit that is indifferent to whether the end user operates under risk neutrality and how they interpret uncertainty in DCA. Still, the bayesDCA R package allows for an easy and comprehensive characterization of uncertainty in DCA, including its expected consequences through EVPI calculation. Judging whether implementation is appropriate given the observed level of uncertainty and context-specific costs depends on both the estimated net benefits and the decision maker's subjective aversion to the potential risk of implementing a suboptimal strategy.

In sum, we propose a method for Bayesian decision curve analysis and provide a freely available implementation in the bayesDCA R package. We hope our contribution will be relevant for studies that involve the validation of clinical prediction models as well as for diagnostic and prognostic test accuracy studies. Ultimately, the Bayesian DCA workflow may help clinicians and health policymakers adopt informed decisions when choosing and implementing clinical decision strategies.

\section{Methods details}
All analyses were performed with R version 4.2.3 within a fixed Docker image \cite{RCoreTeam2022, Merkel2014}. All code and data to reproduce the results are available on GitHub (\textcolor{blue}{\href{https://github.com/giulianonetto/paper-bayesdca}{https://github.com/giulianonetto/paper-bayesdca}}). For the survival case, the bayesDCA R package employs Markov Chain Monte Carlo based on the No U-Turn Sampler implemented within Stan and accessed via the rstan R package \cite{Carpenter2017, rstan2022}. Processing of posterior samples and data visualization employed the tidyverse meta-package and the patchwork package \cite{Wickham2019, Pedersen2022}. Parallel processing for the simulations was implemented using the furrr package \cite{Vaughan2022}. Pipeline management was implemented using the targets package \cite{Landau2021}. The bayesDCA R package is freely available at \textcolor{blue}{\href{https://github.com/giulianonetto/bayesdca}{https://github.com/giulianonetto/bayesdca}}.

\subsection{GUSTO-I trial example}
An illustrative model was built using the GUSTO-I trial dataset\cite{gusto1993}. From the full dataset ($N=40,830$), we held out a randomly selected validation set ($N=500$, $36$ events) and trained a simple logistic regression model on the remaining data based on age, systolic blood pressure, pulse, and Killip class (I -- IV). We ran both Bayesian and Frequentist DCA of the fitted model on the validation data to provide an initial illustration under large uncertainty.
\subsection{Bayesian DCA: model details}
\label{sec:bdca-details}
\subsubsection{Binary outcomes}
We now describe the model used to estimate the net benefit for each decision strategy at each decision threshold. For each patient $i=1, 2, \dots, N$, suppose we observe the pair $(D_i, Z_i)$ where $D_i = 1\{i^{th} \textrm{ patient has disease}\}$ and   $Z_i = 1\{i^{th} \textrm{ patient has positive prediction}\}$. We then model:
\begin{align}
D &\sim \textrm{Bernoulli}(\theta_0)\nonumber\\
Z \vert D =1 &\sim \textrm{Bernoulli}(\theta_1)\\
Z \vert D =0 &\sim \textrm{Bernoulli}(\theta_2)\nonumber
\end{align}
where $\theta_0$ is the prevalence, $\theta_1$ is the sensitivity, and $\theta_2$ is 1 minus the specificity. The likelihood function for the observed data $\mathcal{D}=\big\{\big(D_i, Z_i\big)\big\}_{i=1}^N$ given the parameter vector $\bm{\theta}=(\theta_0\ \theta_1\ \theta_2)$ is:
\begin{align}
L( \mathcal{D}\,\vert\,\bm{\theta}) 
&= 
\prod_{i=1}^N p(d_i, z_i) = \prod_{i=1}^N p(d_i) p(z_i\vert d_i) 
=\prod_{i=1}^N 
    \textrm{Ber}(d_i\vert\theta_0) \times
    \textrm{Ber}(z_i\vert\theta_1)^{d_i} \times
    \textrm{Ber}(z_i\vert\theta_2)^{1-d_i}
\end{align}
The parameters $\theta_0$, $\theta_1$, and $\theta_2$ are orthogonal because the likelihood function factorizes as:
\begin{align}
L( \mathcal{D}\,\vert\,\theta_0, \theta_1, \theta_2) 
&=
\prod_{i=1}^N 
    \textrm{Ber}(d_i\vert\theta_0) \times
\prod_{i=1}^N 
    \textrm{Ber}(z_i\vert\theta_1)^{d_i} \times
\prod_{i=1}^N 
    \textrm{Ber}(z_i\vert\theta_2)^{1-d_i}
\end{align}
Notice that each product term depends on only one parameter. From a Frequentist perspective, this implies that the maximum likelihood estimators for the parameters of interest are asymptotically independent\cite{Cox_1987}. In the Bayesian approach, under independent priors, the factorization above implies posterior independence. In our case, let $\theta_j\sim \textrm{Beta}(\alpha_j, \beta_j)$ for $j=0,1,2$ be our independent priors. The posterior distribution $\pi(\bm{\theta}\vert \mathcal{D})$ is then proportional to:
\begin{align*}
        \Bigg[
        \textrm{Beta}(\alpha_0, \beta_0)
        \prod_{i=1}^N 
            \textrm{Ber}(d_i\vert\theta_0)
        \Bigg]
        \times
        \Bigg[
        \textrm{Beta}(\alpha_1, \beta_1)
        \prod_{i=1}^N 
            \textrm{Ber}(z_i\vert\theta_1)^{d_i}
        \Bigg]
        \times
        \Bigg[
        \textrm{Beta}(\alpha_2, \beta_2)
        \prod_{i=1}^N 
            \textrm{Ber}(z_i\vert\theta_2)^{1-d_i}
        \Bigg]
\end{align*}
which is the numerator of the Bayes' Theorem formula and simplifies to
\begin{align}
    \Bigg[ \theta_0^{D+\alpha_0 - 1}(1-\theta_0)^{ND + \beta_0 - 1}\Bigg]
    \times
    \Bigg[\theta_1^{TP + \alpha_1 - 1}(1-\theta_1)^{FN + \beta_1 -1}\Bigg] 
    \times
    \Bigg[\theta_2^{FP+ \alpha_2 -1}(1-\theta_2)^{TN+\beta_2 - 1}\Bigg]
    \label{post-numerator}
\end{align}
where $D = \sum_id_i$ and $ND=N-D$ are the numbers of patients with and without the disease, respectively, $TP=\sum_{i}d_iz_i$ represents the total number of true positives, $FN=\sum_{i}d_i(1-z_i)$ of false negatives, $FP=\sum_i(1-d_i)z_i$ of false positives, and $TN=\sum_i(1-d_i)(1-z_i)$ of true negatives. The expression above can already be recognized as the joint density of three independent Beta random variables. Now, let $\mathcal{B}(a, b)=\int _{0}^{1}t^{a-1}(1-t)^{b-1}\,dt$ be the Beta function, then:
\begin{align}
    p(\mathcal{D})
    &= 
    \int_0^1\int_0^1\int_0^1
    \pi(\theta_0)\pi(\theta_1)\pi(\theta_2) \times L(\mathcal{D}\,\vert\,\bm{\theta}) \,d\theta_0\,d\theta_1\,d\theta_2\nonumber\\
    &=
    \prod_{j=0}^2 \mathcal{B}(\alpha_j, \beta_j)^{-1}\nonumber\\
    &\ \ \ \ \times
    \int_0^1 \Bigg[ \theta_0^{D+\alpha_0 - 1}(1-\theta_0)^{ND + \beta_0 - 1}\Bigg] \,d\theta_0\nonumber\\
    &\ \ \ \ \times
    \int_0^1 \Bigg[\theta_1^{TP + \alpha_1 - 1}(1-\theta_1)^{FN + \beta_1 -1}\Bigg]  \,d\theta_1
    &\textrm{(from (\ref{post-numerator}))}\nonumber\\
    &\ \ \ \ \times
    \int_0^1 \Bigg[\theta_2^{FP+ \alpha_2 -1}(1-\theta_2)^{TN+\beta_2 - 1}\Bigg]  \,d\theta_2\nonumber\\[10pt]
    &=
    \prod_{j=0}^2 \mathcal{B}(\alpha_j, \beta_j)^{-1}
    \times
    \mathcal{B}(D+\alpha_0, ND+\beta_0)
    \times
    \mathcal{B}(TP+\alpha_1, FN+\beta_1)
    \times
    \mathcal{B}(FP+\alpha_2, TN+\beta_2)\label{post-denominator}
\end{align}
Putting (\ref{post-numerator}) and (\ref{post-denominator}) together and noticing that the normalizing constant that multiplies expression (\ref{post-numerator}) for the posterior distribution is $\Big[\prod_{j=0}^2 \mathcal{B}(\alpha_j, \beta_j)^{-1}\Big]\big/p(\mathcal{D})$, we have by Bayes' Theorem:
\begin{align}
    \pi(\bm{\theta}\vert \mathcal{D}) = 
    \textrm{Beta}(\theta_0 \,\vert\,D+\alpha_0, ND+\beta_0)
    \times
    \textrm{Beta}(\theta_1\,\vert\,TP+\alpha_1, FN+\beta_1)
    \times
    \textrm{Beta}(\theta_2\,\vert\,FP+\alpha_2, TN+\beta_2)
\end{align}
Notice that the joint posterior distribution is the product of the marginal posterior distributions from each parameter -- i.e., posterior independence. Hence, we can simply draw from the marginal posteriors and combine the marginal samples to form a draw from the joint posterior. This result holds for any sample size and not only asymptotically.

Given samples from the joint posterior, we then compute the posterior net benefit as $\textrm{NB}\vert\mathcal{D} = \theta_1 \cdot \theta_0 - w_t \cdot \theta_2 \cdot (1-\theta_0)$, where $w_t = t/(1-t)$. The posterior net benefit for the treat all strategy is given by $\textrm{NB}_{\textrm{all}}\vert\mathcal{D}=\theta_0 - w_t \cdot (1-\theta_0)$. Also, within bayesDCA we sample specificity $\theta_3 = 1 - \theta_2$ directly instead of $\theta_2$ to make communication easier with end users -- in agreement with equation (\ref{bayes-dca-model}) -- and to make prior specification potentially more intuitive. Finally, notice that $\theta_0$ is a common parameter shared by all thresholds and all decision strategies, whereas $\theta_1$ and $\theta_2$ are threshold- and strategy-specific.

\subsubsection{Survival outcomes}
For each patient $i=1, 2, \dots, N$, suppose we observe $(T_i, C_i, Z_i)$ where $T_i$ is the observed survival time for the $i^{th}$ patient, $C_i$ is the censoring indicator,  and   $Z_i = 1\{i^{th} \textrm{ patient has positive prediction}\}$, i.e., $Z_i$ is $1$ if the predicted risk of an event at time horizon $\tau  $ is above the decision threshold. We then model: 
\begin{align}
    Z &\sim \textrm{Bernoulli}(p)\nonumber\\
    (T, C) \,\vert\,Z = 1 &\sim \textrm{Weibull-Censored}(\alpha_1, \sigma_1)\\
    (T, C) \,\vert\,Z = 0 &\sim \textrm{Weibull-Censored}(\alpha_2, \sigma_2)\nonumber\label{dca-weibull}
\end{align}
The likelihood function for the observed data $\mathcal{D} = \big\{(T_i, C_i, Z_i)\big\}_{i=1}^N$
given the parameter vector $\bm{\theta}=\big(p, \alpha_1, \sigma_1, \alpha_2, \sigma_2\big)$ is:
\begin{align}
L\Big(\mathcal{D}\,\vert\,\bm{\theta}  \Big) 
&= \prod_{i=1}^N\textrm{W-Cens}(t_i, c_i \,\vert\,\alpha_1, \sigma_1)^{z_i} \times \textrm{W-Cens}(t_i, c_i \,\vert\,\alpha_2, \sigma_2)^{1-z_i} \times \textrm{Bern}(z_i \,\vert\,p)\nonumber\\
&=\prod_{i=1}^N\textrm{W-Cens}(t_i, c_i \,\vert\,\alpha_1, \sigma_1)^{z_i} \times \prod_{i=1}^N\textrm{W-Cens}(t_i, c_i \,\vert\,\alpha_2, \sigma_2)^{1-z_i} \times \prod_{i=1}^N \textrm{Bern}(z_i \,\vert\,p)\\
&:=
L_1\big(\mathcal{D}_+\,\vert\,\bm{\theta_1}\big)
\times
L_2\big(\mathcal{D}_-\,\vert\,\bm{\theta_2}\big)
\times
L_3\big(\mathcal{D}_0\,\vert\,p\big)
\end{align}
where we represent the data as $\mathcal{D}_+ = \big\{T_i, C_i\big\}_{i\in [n]: z_i=1}$ (survival data for patients with positive predictions), $\mathcal{D}_- = \big\{T_i, C_i\big\}_{i\in [n]: z_i=0}$ (survival under negative predictions), $\mathcal{D}_0 = \big\{Z_i\big\}_{i=1}^n$ (positive prediction indicators), and Weibull parameters as $\bm{\theta_1}=(\alpha_1, \sigma_1)$ and $\bm{\theta_2}=(\alpha_2, \sigma_2)$. Hence, under proper independent priors:
\begin{align}
    \pi(\bm{\theta}\,\vert\,\mathcal{D}) \propto
    \pi(\bm{\theta_1})L_1\big(\mathcal{D}_+\,\vert\,\bm{\theta_1}\big)
    \times
    \pi(\bm{\theta_2})L_2\big(\mathcal{D}_-\,\vert\,\bm{\theta_2}\big)
    \times
    \pi(p)L_1\big(\mathcal{D}_0\,\vert\,p\big)
\end{align}
which then implies posterior independence and we can sample each parameter independently from their marginal posteriors. In particular, we place a conjugate Beta(a, b) prior on the Bernoulli parameter $p$ so that we have a closed-form solution for its marginal posterior -- default being $a=1$ and $b=1$. As there is no such closed form for the Weibull likelihood under right-censoring, we need to employ MCMC to sample from the marginal posterior of $\bm{\theta_1}$. Within bayesDCA, this is implemented using Stan to sample $\bm{\theta_1}$ as a parameter and $p$ as a generated quantity\cite{Carpenter2017}. The default priors for $\bm{\theta_1}$ are as in (\ref{surv-priors}), though Gamma priors are also allowed. Due to parameter orthogonality and posterior independence, we don't need to sample the nuisance parameter $\bm{\theta_2}$. We can then compute the posterior conditional survival at the time horizon $\tau$ as $S=\exp\big(\tau/\sigma_1\big)^{\alpha_1}$ and the posterior net benefit as $\textrm{NB}\,\vert\,\mathcal{D}=\big(1-S\big)\cdot p - w_t\cdot S \cdot p$. Notice that here both $S$ and $p$ are threshold- and strategy-specific. For the treat-all strategy, $p$ is a fixed number at 1.
\subsection{Simulation details}
\label{sec:sim-details}
For each simulation setting, we simulate a large population dataset ($N=2\times 10^6$) from which we randomly generate reasonably-sized samples to perform DCA. We set an expected value of 100 events for the simulated samples and varied the sample size according to each setting's outcome prevalence or incidence. Bayesian DCA was compared with the Frequentist counterpart using available open-source software, including the packages rmda and dcurves \cite{rmda2018, dcurves2021}.
\subsubsection{Binary outcomes}
For the binary outcome simulations, the underlying data-generating process is as follows:
\begin{align*}
    &y_i \sim \textrm{Bernoulli}(p_i) 
    \ \  \ \ \textrm{for $i = 1, 2, \dots, N$}\\
    &p_i = \textrm{logit}^{-1}(\bm{z}_i^T \bm{\beta})
    \ ,\ \ \ \ \ 
    \widehat{p}_i = \textrm{logit}^{-1}(\bm{z}_i^T \widehat{\bm{\beta}})\\
    &\bm{z}_i^T=\begin{bmatrix}1&x_{i1}&x_{i2}\end{bmatrix}
    \ ,\ \ \ \ \ x_{i1},\ x_{i2} \overset{\textrm{iid}}{\sim} \textrm{Exp}(1)
\end{align*}
where $\beta$ is a vector of true coefficients used to generate the data, and $\widehat{\beta}$ is a vector of ``estimated'' coefficients from the model under investigation -- the model we are validating with DCA. The true underlying disease probabilities are represented by $p_i$, and the estimated probabilities are $\widehat{p}_i$. We choose values of $\beta$ and $\widehat{\beta}$ to yield settings with a range of values for outcome prevalence, signal-to-noise ratio, and model discrimination. The signal-to-noise ratio is represented by the maximum possible AUC any model could achieve in a given setting (i.e., the AUC computed using the true disease probabilities $p_i$). The discrimination of the model under investigation is the true AUC computed using $\widehat{p}_i$ in the entire population data.

Here, we choose $\bm{\beta}$ and $\widehat{\bm{\beta}}$ to fix the true prevalence at 1\%, 5\%, or 30\%, and the maximum AUC at either 0.65 or 0.85 -- these represent low and high signal-to-noise ratio settings, respectively, with varying prevalence. We choose $\widehat{\bm{\beta}}$ so that the example model approximates the maximum AUC in a given setting as well as the true underlying prevalence, but with poor calibration slope (i.e., the coefficients for $x_{i1}$ and $x_{i2}$ are exaggerated). This resembles a common scenario of overfitting in risk prediction. The fixed parameters for each simulation setting are provided in Table \ref{tab:simulation-settings}.

\subsubsection{Survival outcomes}
For the survival outcome simulations, we employed the simsurv package\cite{Brilleman_2021}. Briefly, we sample survival times from a Weibull distribution with specified shape and scale parameters as well as two standard normal covariates with fixed coefficients (under proportional hazards assumption). Then, we sample censoring times from a uniform distribution with support between zero and 24 months, representing a maximum follow-up time of two years. For each patient in the population, the observed time is the minimum between survival and censoring times. Table (\ref{tab:simulation-settings-surv}) shows the fixed parameters for each simulation setting. Notice that simsurv employs a different Weibull parameterization than the one used by Stan and model (\ref{dca-weibull}); while the shape parameter ($\gamma$ in simsurv) is the same, the scale is denoted $\lambda=\sigma^{-\alpha}$. Table (\ref{tab:simulation-settings-surv}) uses simsurv notation for easy reproduction.

\begin{table}[ht]
\caption{\textbf{Simulation settings (binary outcomes).} AUC refers to the maximum AUC possible in a given setting and Prev. is the true underlying prevalence. The regression coefficients used to generate the data are $\bm{\beta}$, while $\widehat{\bm{\beta}}$ define the hypothetical models under validation with DCA.}
\begin{tabular*}{\linewidth}{cccc}
\toprule
\textbf{AUC} & \multicolumn{1}{c}{\textbf{Prev.}} & \multicolumn{1}{c}{$\bm{\beta}^T$} & \multicolumn{1}{c}{$\widehat{\bm{\beta}}^T$} \\ \midrule
0.65 & 1\% &
$\begin{pmatrix}-4.750&-\log\big[1.50\big]&\log\big[1.50\big]\end{pmatrix}$&
$\begin{pmatrix}-5.000&-\log\big[1.50\big]*1.25&\log\big[1.50\big]*1.25\end{pmatrix}$\\[7pt]
0.65 & 5\% &
$\begin{pmatrix}-3.100&-\log\big[1.50\big]&\log\big[1.50\big]\end{pmatrix}$&
$\begin{pmatrix}-3.900&-\log\big[1.50\big]*3.00&\log\big[1.50\big]*3.00\end{pmatrix}$\\[7pt]
0.65 & 30\% &
$\begin{pmatrix}-0.900&-\log\big[1.55\big]&\log\big[1.55\big]\end{pmatrix}$&
$\begin{pmatrix}-1.200&-\log\big[1.55\big]*3.00&\log\big[1.55\big]*3.00\end{pmatrix}$\\[7pt]
0.85 & 1\% &
$\begin{pmatrix}-5.600&-\log\big[2.57\big]&\log\big[2.57\big]\end{pmatrix}$&
$\begin{pmatrix}-6.900&-\log\big[2.57\big]*1.50&\log\big[2.57\big]*1.50\end{pmatrix}$\\[7pt]
0.85 & 5\% &
$\begin{pmatrix}-3.755&-\log\big[2.95\big]&\log\big[2.95\big]\end{pmatrix}$&
$\begin{pmatrix}-7.300&-\log\big[2.95\big]*3.00&\log\big[2.95\big]*3.00\end{pmatrix}$\\[7pt]
0.85 & 30\% &
$\begin{pmatrix}-1.300&-\log\big[4.50\big]&\log\big[4.50\big]\end{pmatrix}$&
$\begin{pmatrix}-2.250&-\log\big[4.50\big]*3.00&\log\big[4.50\big]*3.00\end{pmatrix}$\\
\bottomrule
\end{tabular*}
\label{tab:simulation-settings}
\end{table}

\begin{table}[ht]
\caption{\textbf{Simulation settings (survival outcomes).} \textbf{C} refers to the maximum C-statistic possible in a given setting, and $S(1)$ is the true underlying one-year survival rate. The $\gamma$ and $\lambda$ parameters match the definitions used by the simsurv R package. The regression coefficients used to generate the data are $\bm{\beta}$, while $\widehat{\bm{\beta}}$ define the hypothetical models under validation with DCA.}
\begin{tabular*}{\linewidth}{cccccc}
\toprule
\textbf{C} & \multicolumn{1}{c}{$S(1)$} & \multicolumn{1}{c}{$\gamma$ (shape)} & \multicolumn{1}{c}{$\lambda$ (scale)} & \multicolumn{1}{c}{$\bm{\beta}^T$} & \multicolumn{1}{c}{$\widehat{\bm{\beta}}^T$} \\ \midrule
0.60 & 10\% &1.22&0.12&
$\begin{pmatrix}\log\big[1.30\big]&\log\big[0.70\big]\end{pmatrix}$&
$\begin{pmatrix}\log\big[1.30\big]&\log\big[0.70\big]\end{pmatrix}*1.01$\\[7pt]
0.60 & 20\% &1.07&0.12&
$\begin{pmatrix}\log\big[1.30\big]&\log\big[0.70\big]\end{pmatrix}$&
$\begin{pmatrix}\log\big[1.30\big]&\log\big[0.70\big]\end{pmatrix}*1.01$\\[7pt]
0.60 & 50\% &0.7&0.12&
$\begin{pmatrix}\log\big[1.30\big]&\log\big[0.70\big]\end{pmatrix}$&
$\begin{pmatrix}\log\big[1.30\big]&\log\big[0.70\big]\end{pmatrix}*1.01$\\[7pt]
0.90 & 10\% &4.60&0.0004&
$\begin{pmatrix}\log\big[1.95\big]&\log\big[0.05\big]\end{pmatrix}$&
$\begin{pmatrix}\log\big[1.95\big]&\log\big[0.05\big]\end{pmatrix}*1.25$\\[7pt]
0.90 & 20\% &4.00&0.0004&
$\begin{pmatrix}\log\big[1.95\big]&\log\big[0.05\big]\end{pmatrix}$&
$\begin{pmatrix}\log\big[1.95\big]&\log\big[0.05\big]\end{pmatrix}*1.25$\\[7pt]
0.90 & 50\% &2.90&0.0004&
$\begin{pmatrix}\log\big[1.95\big]&\log\big[0.05\big]\end{pmatrix}$&
$\begin{pmatrix}\log\big[1.95\big]&\log\big[0.05\big]\end{pmatrix}*1.25$\\[7pt]
0.95 & 10\% &6.50&0.0004&
$\begin{pmatrix}\log\big[1.95\big]&\log\big[0.001\big]\end{pmatrix}$&
$\begin{pmatrix}\log\big[1.95\big]&\log\big[0.001\big]\end{pmatrix}*1.25$\\[7pt]
0.95 & 20\% &5.40&0.0004&
$\begin{pmatrix}\log\big[1.95\big]&\log\big[0.001\big]\end{pmatrix}$&
$\begin{pmatrix}\log\big[1.95\big]&\log\big[0.001\big]\end{pmatrix}*1.25$\\[7pt]
0.95 & 50\% &3.10&0.0004&
$\begin{pmatrix}\log\big[1.95\big]&\log\big[0.001\big]\end{pmatrix}$&
$\begin{pmatrix}\log\big[1.95\big]&\log\big[0.001\big]\end{pmatrix}*1.25$\\[7pt]
\bottomrule
\end{tabular*}
\label{tab:simulation-settings-surv}
\end{table}

\section{Acknowledgements}
The authors thank Dr. Mohsen Sadatsafavi for extensive feedback and suggestions on drafts of this manuscript, as well as all members of the Korthauer lab for helpful comments. We also thank Dr. Andrew Vickers and Dr. Paul Gustafson for insightful discussion of the work. We also gratefully acknowledge the funding support from the BC Children's Hospital Research Institute Establishment Award (to KK).
\newpage
\bibliographystyle{ieeetr}
\bibliography{getwriting}
\clearpage
\setcounter{figure}{0}    
\renewcommand{\figurename}{Supplementary Figure }
\renewcommand{\thefigure}{S\arabic{figure}}
\setcounter{table}{0}    
\renewcommand{\tablename}{Supplementary Table}
\renewcommand{\thetable}{S\arabic{table}}
\makeatletter
\def\fnum@figure{\figurename \thefigure}
\makeatother
\renewcommand{\thesection}{}
\renewcommand{\thesubsection}{}
\makeatletter
\def\@seccntformat#1{\csname the#1\endcsname}
\makeatother
\section{Supplement}
\subsection{Supplementary methods}
\makeatletter
\def\@seccntformat#1{\csname the#1\endcsname\quad}
\makeatother
\renewcommand{\thesubsubsection}{\arabic{subsubsection}}
\subsubsection{Informative priors in Bayesian DCA for binary outcomes}
\label{supp:inf-priors}
In this section, we briefly describe how informative priors can be easily employed within bayesDCA for binary outcomes. These can be advantageous under small effective sample sizes, for instance, for the estimation of EVPI\cite{Sadatsafavi2023}. Additionally, we demonstrate the visual prior predictive checks available within bayesDCA to aid in understanding the implications of the chosen prior parameters\cite{Gelman2020}.

In the proposed Bayesian DCA approach, each parameter of interest (sensitivity, specificity, and prevalence) follows an independent Beta distribution -- a priori and a posteriori. For interpretability, we parameterize the Beta priors using the prior mean and prior sample size (or ``prior strength"). For instance, suppose we state a priori that the prevalence $p$ follows a Beta distribution with mean $\mu$ and sample size (or strength) $\eta$. Then $p\sim \textrm{Beta}(\mu\,\eta ,\ (1-\mu)\,\eta)$ a priori.

In DCA for binary outcomes, sensitivity (Se) decreases as the decision threshold increases, and the reverse is true for specificity. This prior knowledge motivates a threshold-varying prior implemented within bayesDCA as follows: for a given decision strategy (e.g., a diagnostic model), let $\textrm{Se}_t$ be the (prior) sensitivity at the decision threshold $t$. Define an ``ignorance region" $R=(a, b)$ with $0\leq t_{\min}<a<b<t_{\max}<1$ where we have no prior knowledge about the sensitivity of the decision strategy. Here $t_{\min}$ is the lowest decision threshold considered in the DCA (usually 0) and $t_{\max}$ the highest. At $t_{\min}$, we strongly believe the sensitivity is high, which translates into a high prior mean and a large prior sample size. As the decision threshold increases, the sensitivity should decrease, though by an uncertain amount. Hence, we progressively decrease both the prior mean and prior sample size. As the decision threshold approaches $a$, the prior mean approaches $0.5$ and the prior sample size approaches $2$, which together represent the vague $\textrm{Beta}(1,1)$ prior -- fixed for any $t$ between $a$ and $b$. The prior mean starts decreasing again for $t>b$, reaching its minimum at $t_{\max}$, whereas the prior sample size now increases (at the same rate it previously decreased for $t<a$). Since we expect an opposite behaviour for the specificity, we set its prior mean to be the opposite of the prior mean sensitivity, for any given threshold. Supplementary Figure (\ref{fig:ppc}) shows the resulting prior.

When using this threshold-varying prior, the default ignorance region $R$ within bayesDCA is $(0.25,\ 0.75) \times t_{\max}$. Optionally, the user may specify prior mean and sample size other than 0.5 and 2, respectively, to be fixed within this region. The default prior mean sensitivities are 0.99 and 0.01 at $t_{\min}$ and $t_{\max}$, respectively. The default prior sample size for sensitivity and specificity is 5 at both $t_{\min}$ and $t_{\max}$ and 2 within the ignorance region. We employ linear interpolation to get prior means and sample sizes at decision thresholds between $t_{\min}$ an $a$ and between $b$ and $t_{\max}$. We remain vague about the prevalence by default as it is highly context-dependent, but users should set more reasonable parameters for their problem.

The threshold-varying prior above is not particularly informative, in the sense that it can hardly impact the net benefit estimates for reasonably large datasets (e.g., with 100 events or more). Yet, it is a starting point from which we can build more informative priors that make sense in any given context. For instance, suppose we wish to set the prior sensitivity to linearly decrease with the decision threshold, without any ignorance region; we also believe that the prevalence is likely between 20\% and 40\%, with the best guess at 30\%. With bayesDCA, it is easy to employ such priors and visualize its implications, shown in Supplementary Figure (\ref{fig:ppc2}).

Notice that, because we ``turned off" the ignorance region, the default prior sample size for sensitivity and specificity is fixed at 5 (default). Still, there is considerable prior uncertainty for these parameters depending on the decision threshold. The prior on the prevalence parameter completely determines the prior net benefit for the Treat all strategy (recall that $NB_{\textrm{all}} = p - w_t\,(1-p)$ where $w_t=t/(1-t)$). However, the prior net benefit of a hypothetical model-based strategy (to be assessed using DCA) closely follows the Treat all strategy for many thresholds. Even though our prior sample size for sensitivity and specificity is not large, the prior prevalence has a particularly strong effect on the implied prior net benefit for all decision strategies. In particular, it sets an upper bound on the prior net benefit -- here, around 0.4. In this example, the effect of the prior sensitivity and specificity is more pronounced at higher thresholds, where the prior net benefit from the Treat all and model-based strategies start to diverge more notably.

Though the user of bayesDCA may employ arbitrary prior parameters if desired, the above examples illustrate what is easily accessible in terms of threshold-varying priors within the package. The ignorance region enables prior vagueness for an arbitrary portion of the decision curve while still taking advantage of prior knowledge about sensitivity and specificity at very high and very low thresholds. The prior prevalence has en effect of restricting the range of prior net benefit that is plausible for any given sensitivity and specificity. Future work may further develop prior distributions based on prior knowledge about discrimination and calibration as well as on ideas from signal detection theory\cite{Sadatsafavi_2007}. Finally, the implications of any prior choices can be visualized with prior predictive checks.

\subsubsection{Informative priors preserve EVPI monotonic behaviour}
\label{supp:evpi-monotonic}
To show how we can leverage informative priors to preserve the expected monotonic behaviour of the EVPI, we adapt the simulation code published in Sadatsafavi et al. (2023)\cite{Sadatsafavi2023}. Briefly, using an example model trained with a subset of the GUSTO-I trial data, we run DCA and EVPI calculations in an external dataset of increasing sample size (held out GUSTO-I data) as previously described\cite{Sadatsafavi2023}. We repeat this procedure for four decision thresholds: 0.01, 0.02, 0.05, and 0.1. Previously, Sadatsafavi et al. (2023)\cite{Sadatsafavi2023} identified unexpected behaviour of EVPI for the 0.01 threshold: the EVPI would increase and then decrease with sample size (see Figure 5 in the publication).

Here, we repeat their simulation but adding two methods: bayesDCA with uniform $\textrm{Beta}(1, 1)$ priors on all parameters (``BayesDCA (uniform)") and bayesDCA with informative priors (``BayesDCA (informative)"). For the informative case, we employed bayesDCA's framework for threshold-varying priors described in the previous section (using the .get\_prior\_parameters() function implemented in bayesDCA). A summary of the resulting prior distributions for sensitivity and specificity is shown in Supplementary Table (\ref{tab:evpi-simulation}). The prior sample size was fixed at 10 and prevalence used a uniform Beta(1, 1) prior. Supplementary Figure (\ref{fig:evpi-simulation}) shows the results.

\smallskip

\begin{table}[ht]
\caption{\textbf{Summary of informative Beta priors used in EVPI simulation.}}
\begin{tabular*}{\linewidth}{ccccccccc}
\hline\noalign{\smallskip}
 & \multicolumn{4}{c}{Sensitivity} & \multicolumn{4}{c}{Specificity}\\[5pt]

 \multicolumn{1}{l|}{Threshold}& \multicolumn{1}{c}{Mean} & \multicolumn{1}{c}{95\% Cr.I.} & \multicolumn{1}{c}{Shape 1} & \multicolumn{1}{c|}{Shape 2} & \multicolumn{1}{c}{Mean} & \multicolumn{1}{c}{95\% Cr.I.} & \multicolumn{1}{c}{Shape 1} & \multicolumn{1}{c}{Shape 2} \\

 \multicolumn{1}{l|}{0.01} & \multicolumn{1}{c}{0.95}  & \multicolumn{1}{c}{(0.76\,---\,1.00)} & \multicolumn{1}{c}{9.5} & \multicolumn{1}{c|}{0.5} & \multicolumn{1}{c}{0.05} & \multicolumn{1}{c}{(0.00\,---\,0.24)} & \multicolumn{1}{c}{0.5} & \multicolumn{1}{c}{9.5}\\

 \multicolumn{1}{l|}{0.02} & \multicolumn{1}{c}{0.90} & \multicolumn{1}{c}{(0.66\,---\,1.00)} & \multicolumn{1}{c}{9.0} & \multicolumn{1}{c|}{1.0} & \multicolumn{1}{c}{0.10}  & \multicolumn{1}{c}{(0.00\,---\,0.34)} & \multicolumn{1}{c}{1.0} & \multicolumn{1}{c}{9.0} \\

 \multicolumn{1}{l|}{0.05} & \multicolumn{1}{c}{0.75} & \multicolumn{1}{c}{(0.46\,---\,0.95)} & \multicolumn{1}{c}{7.5} & \multicolumn{1}{c|}{2.5} & \multicolumn{1}{c}{0.25} & \multicolumn{1}{c}{(0.05\,---\,0.54)} & \multicolumn{1}{c}{2.5} & \multicolumn{1}{c}{7.5}\\

 \multicolumn{1}{l|}{0.1} & \multicolumn{1}{c}{0.50} & \multicolumn{1}{c}{(0.21\,---\,0.79)} & \multicolumn{1}{c}{5.0} & \multicolumn{1}{c|}{5.0} & \multicolumn{1}{c}{0.50} & \multicolumn{1}{c}{(0.21\,---\,0.79)} & \multicolumn{1}{c}{5.0} & \multicolumn{1}{c}{5.0}\\

 \hline 

\multicolumn{9}{l}{\small*Prior sample size was fixed at 10.}\\
\multicolumn{9}{l}{\small**A uniform Beta(1, 1) prior was used for prevalence.}
\end{tabular*}
\label{tab:evpi-simulation}
\end{table}

For all methods except bayesDCA with informative priors, the non-monotonic behaviour is observed at the lowest decision threshold of 0.01. The informative prior recovers the expected monotonic behaviour, adequately indicating higher EVPI for lower sample sizes in all decision thresholds. Thus, bayesDCA offers the ability to leverage prior information to improve EVPI calculations in settings with small effective sample sizes.

\clearpage

\makeatletter
\def\@seccntformat#1{\csname the#1\endcsname}
\makeatother
\subsection{Supplementary Figures}
\makeatletter
\def\@seccntformat#1{\csname the#1\endcsname\quad}
\makeatother

\begin{figure}[H]
\captionsetup{width=.95\linewidth}
\begin{center}
\includegraphics[width=.95\linewidth]{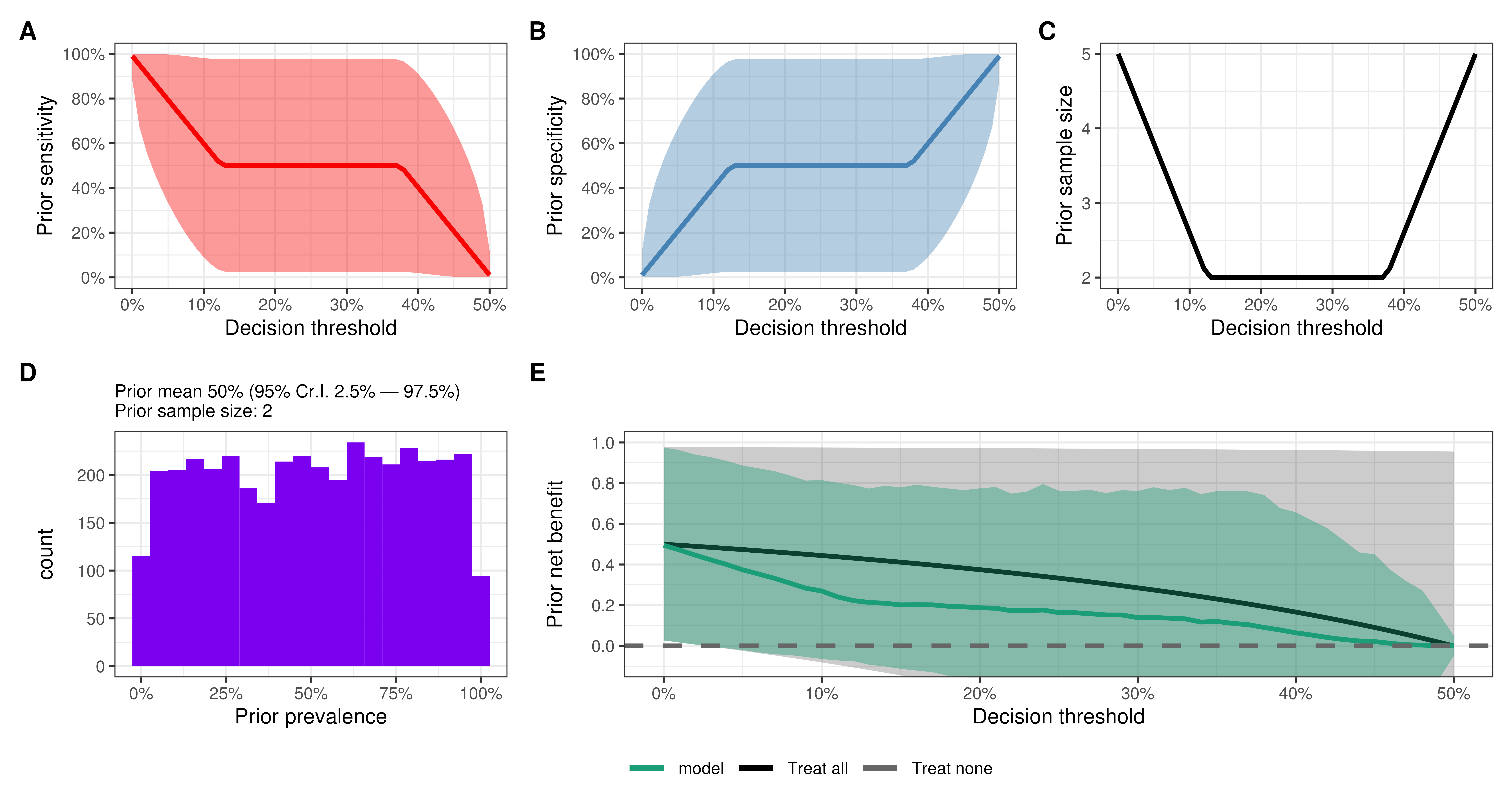}
\end{center}
\caption{\textbf{bayesDCA allows easy implementation and visualization of threshold-varying priors.} \textbf{(A)} The prior mean and prior ``strength" for sensitivity start high at the lowest decision threshold and decrease linearly until reaching a region of ignorance, where the prior sensitivity is Beta(1, 1) (i.e., uniform). After the region of ignorance, the prior mean starts decreasing again, but the prior strength increases. \textbf{(B)} The prior specificity shows the exact opposite behaviour as the prior sensitivity. \textbf{(C)} The prior sample size for both sensitivity and specificity defines the prior strength, is high for extreme decision thresholds (i.e., near the minimum and maximum thresholds), and low within the region of ignorance. \textbf{(D)} The default prior prevalence is the vague Beta(1,1) as it is highly context-dependent. \textbf{(E)} The implications of these priors on the prior net benefit for the Treat all strategy and a hypothetical model-based strategy. By default, the same threshold-varying prior is applied to all decision strategies being analyzed.}
\label{fig:ppc}
\end{figure}

\begin{figure}[H]
\captionsetup{width=.95\linewidth}
\begin{center}
\includegraphics[width=.95\linewidth]{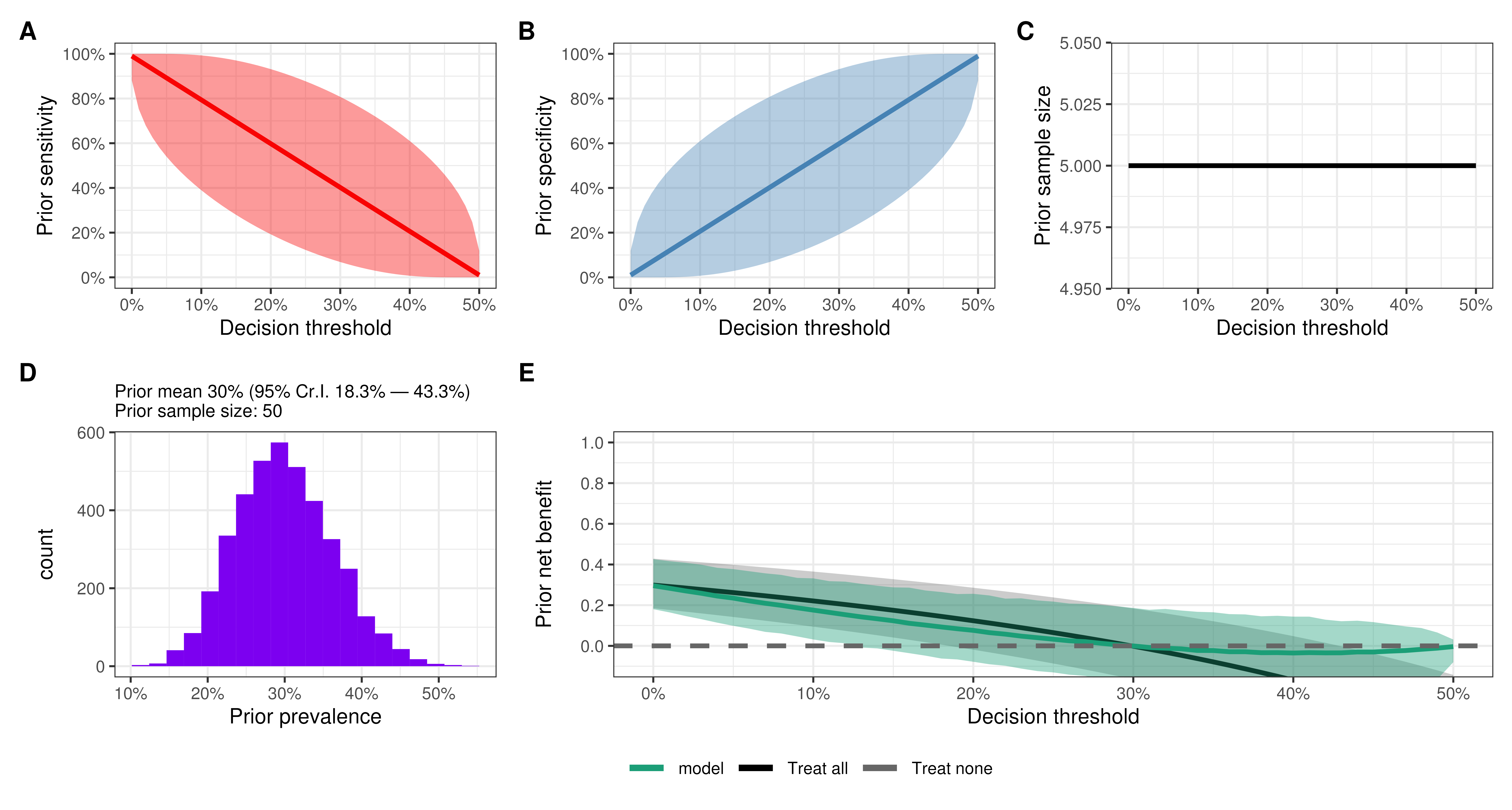}
\end{center}
\caption{\textbf{bayesDCA allows easy implementation and visualization of strongly informative threshold-varying priors.} \textbf{(A)} The prior mean sensitivity decreases linearly with the decision threshold. Although prior ``strength" is fixed, the prior uncertainty varies and is highest when the prior mean is 50\%. \textbf{(B)} The prior specificity shows the exact opposite behaviour as the prior sensitivity. \textbf{(C)} The prior sample size for both sensitivity and specificity is fixed in this case. \textbf{(D)} A prior prevalence suggesting the true prevalence is likely between 20\% and 40\%, with best guess at 30\%. \textbf{(E)} The implications of these priors on the prior net benefit for the Treat all strategy and a hypothetical model-based strategy. The prior prevalence restricts the prior net benefit to be mostly below 0.4, and the priors on sensitivity and specificity drive any differences between the Treat all and model-based decision strategies.}
\label{fig:ppc2}
\end{figure}

\begin{figure}[ht]
\begin{center}
\includegraphics[width=.95\linewidth]{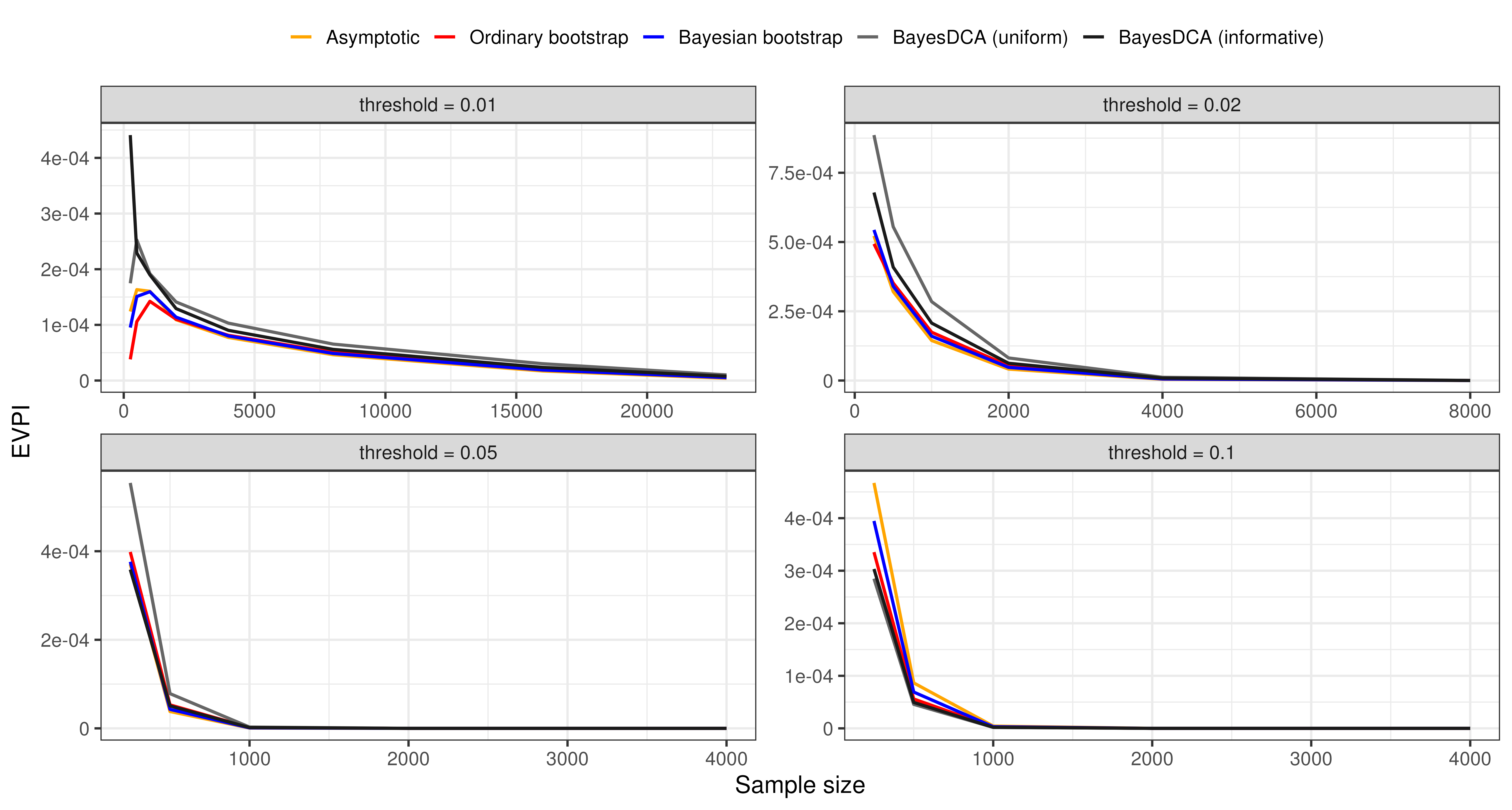}
\end{center}
\caption{\textbf{Informative priors preserve EVPI monotonic behaviour.} EVPI simulation adapted from Sadatsafavi et al. (2023)\cite{Sadatsafavi2023} using GUSTO-I trial data as an example (see Figure 5 in the cited manuscript). Each panel is one decision threshold, and the validation sample size is shown on the x-axis. Informative prior parameters for bayesDCA are reported in Table (\ref{tab:evpi-simulation}).}
\label{fig:evpi-simulation}
\end{figure}

\begin{figure}[H]
\captionsetup{width=.95\linewidth}
\begin{center}
\includegraphics[width=.95\linewidth]{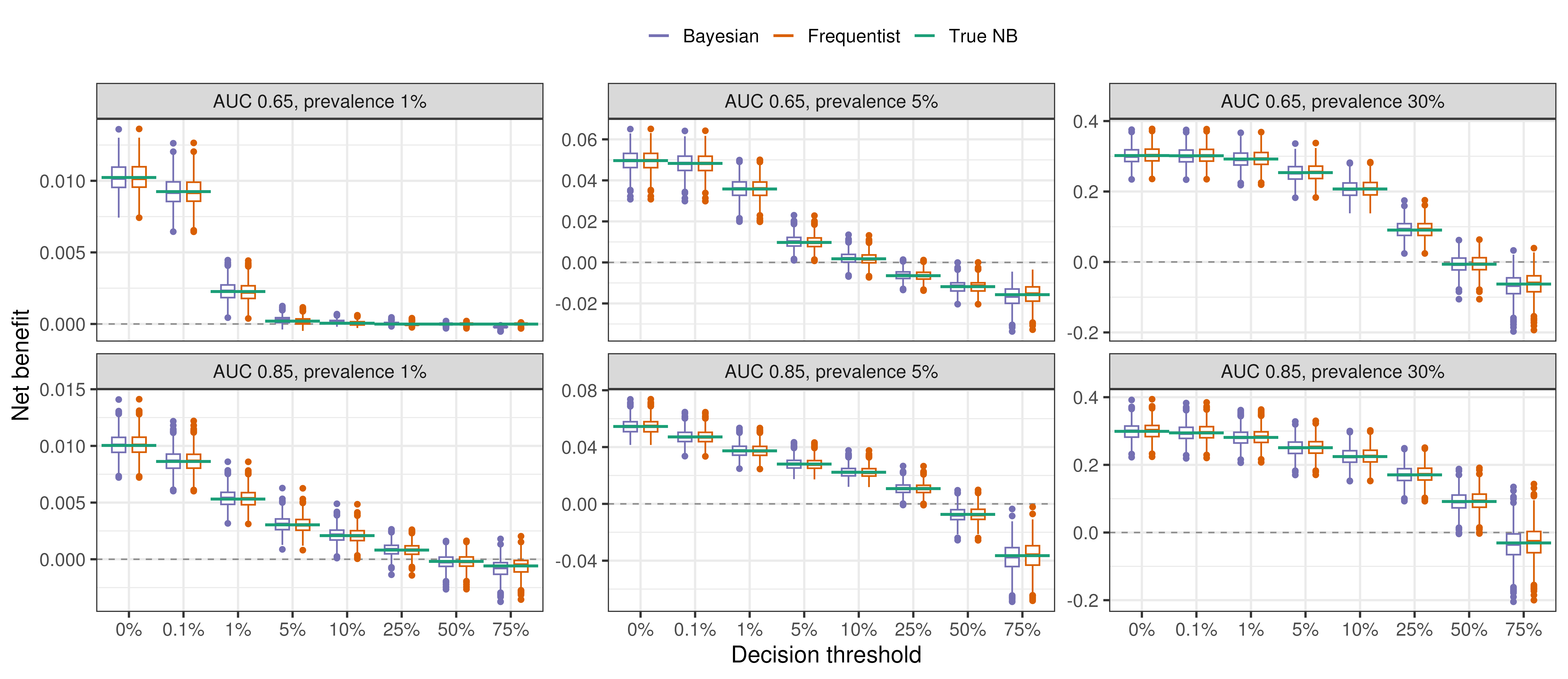}
\end{center}
\caption{\textbf{Bayesian and Frequentist DCA for binary outcomes show similar point estimate distributions.} Bayesian DCA was computed using the bayesDCA R package, while the Frequentist alternative used the bootstrap-based rmda package. For each simulation run, DCA was performed for a fixed example model using a simulated test dataset of sample size corresponding to 100 expected events. A total of \simruns Monte Carlo repetitions was run for each setting. The setting AUC corresponds to its maximum achievable AUC. The example model for each setting was fixed to approximate the maximum discrimination of that setting but was miscalibrated (overly extreme risk predictions).}
\label{fig:point-estimates-distribution}
\end{figure}

\begin{figure}[H]
\captionsetup{width=.95\linewidth}
\begin{center}
\includegraphics[width=.95\linewidth]{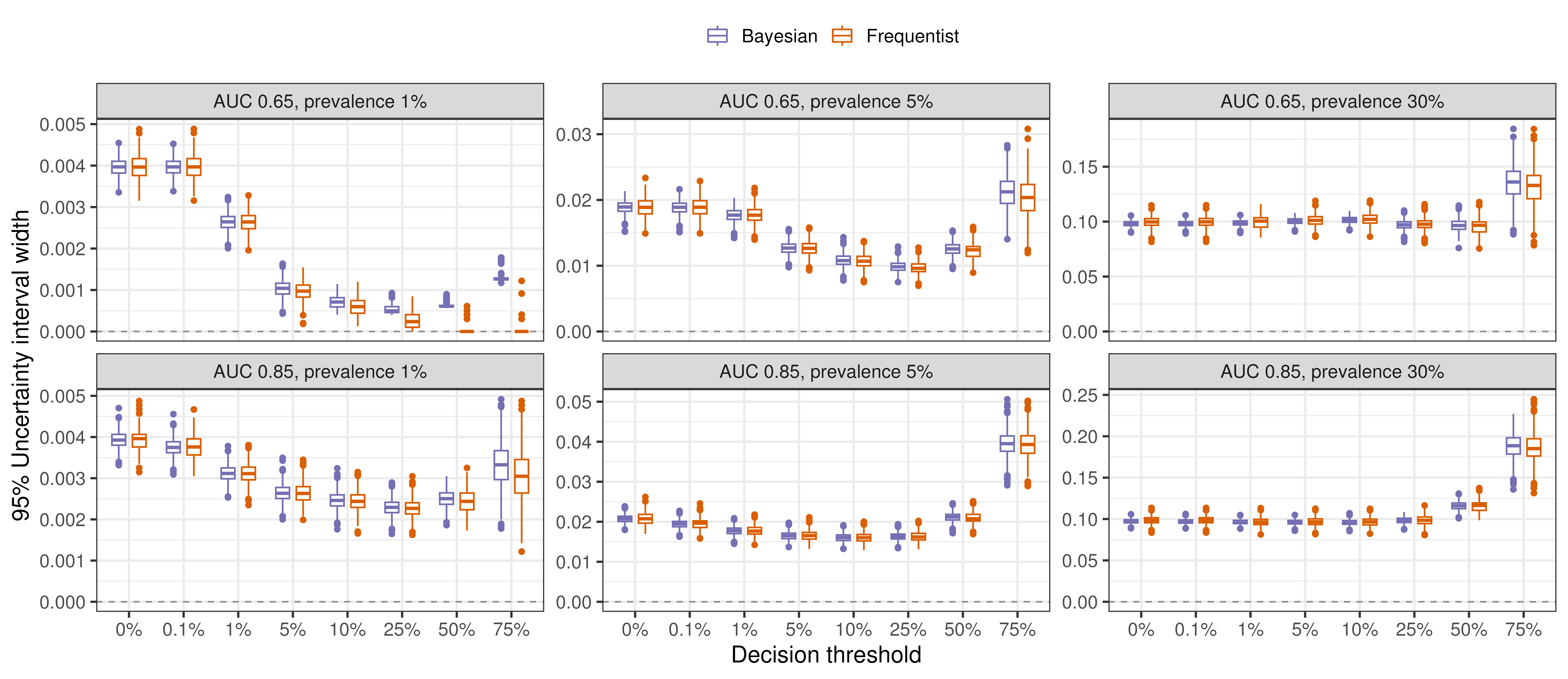}
\end{center}
\caption{\textbf{Bayesian and Frequentist DCA for binary outcomes show similar width of uncertainty intervals, except when bootstrap fails.} Bayesian DCA was computed using the bayesDCA R package, while the Frequentist alternative used the bootstrap-based rmda package. For each simulation run, DCA was performed for a fixed example model using a simulated test dataset of sample size corresponding to 100 expected events. A total of \simruns Monte Carlo repetitions was run for each setting. The setting AUC corresponds to its maximum achievable AUC. The example model for each setting was fixed to approximate the maximum discrimination of that setting but was miscalibrated (overly extreme risk predictions).}
\label{fig:ci-width}
\end{figure}

\begin{figure}[H]
\captionsetup{width=.95\linewidth}
\begin{center}
\includegraphics[width=.95\linewidth]{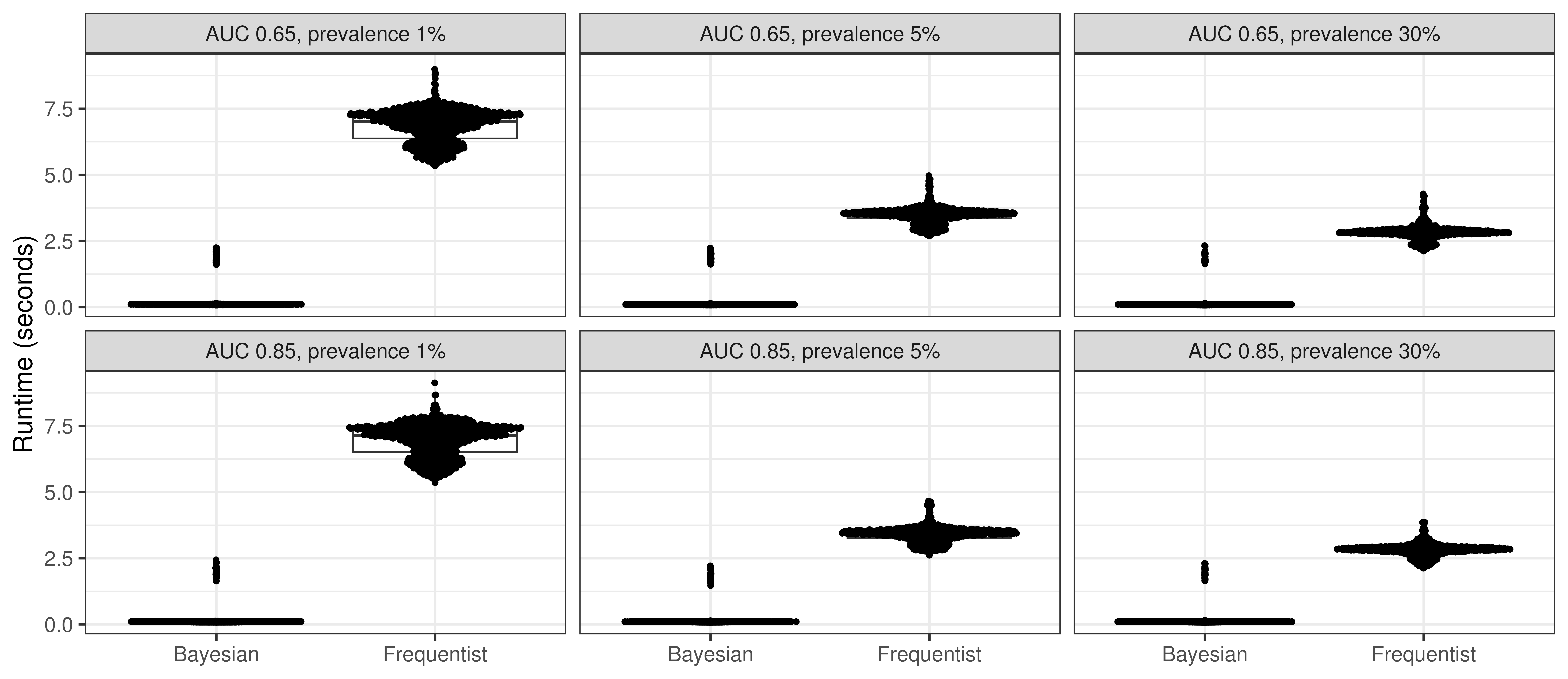}
\end{center}
\caption{\textbf{Bayesian DCA for binary outcomes is faster than bootstrap-based Frequentist DCA.} Bayesian DCA was computed using the bayesDCA R package, while the Frequentist alternative used the bootstrap-based rmda package. For each simulation run, DCA was performed for a fixed example model using a simulated test dataset of sample size corresponding to 100 expected events. A total of \simruns Monte Carlo repetitions was run for each setting. The setting AUC corresponds to its maximum achievable AUC. The example model for each setting was fixed to approximate the maximum discrimination of that setting but was miscalibrated (overly extreme risk predictions). Computation time varies significantly with the overall sample size (approximately 100/\textrm{prevalence}) for the Frequentist case, but not in the Bayesian case.}
\label{fig:runtime}
\end{figure}

\begin{figure}[H]
\captionsetup{width=.95\linewidth}
\begin{center}
\includegraphics[width=.95\linewidth]{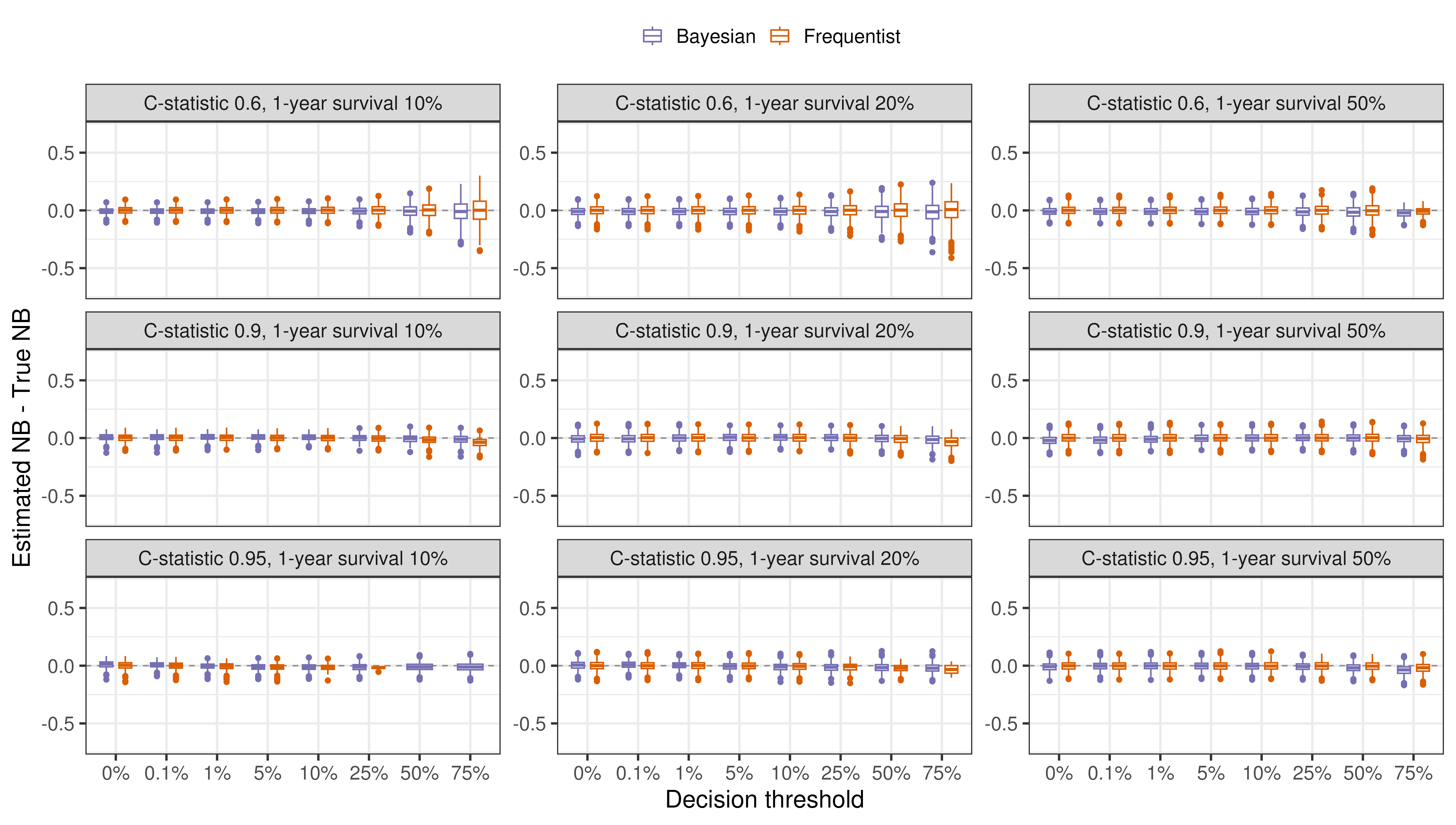}
\end{center}
\caption{\textbf{Bayesian and Frequentist DCA for survival outcomes show similar distributions of point estimate errors.} Bayesian DCA was computed using the bayesDCA R package, while the Frequentist alternative used the dcurves package. For each simulation run, DCA was performed for a fixed example model using a simulated test dataset of sample size corresponding to 100 expected events. A total of \simruns Monte Carlo repetitions was run for each setting. The setting C-statistic corresponds to its maximum achievable C-statistic. The example model for each setting was fixed to approximate the maximum discrimination of that setting but was miscalibrated (overly extreme risk predictions).}
\label{fig:point-estimates-error-surv}
\end{figure}

\begin{figure}[H]
\captionsetup{width=.95\linewidth}
\begin{center}
\includegraphics[width=.95\linewidth]{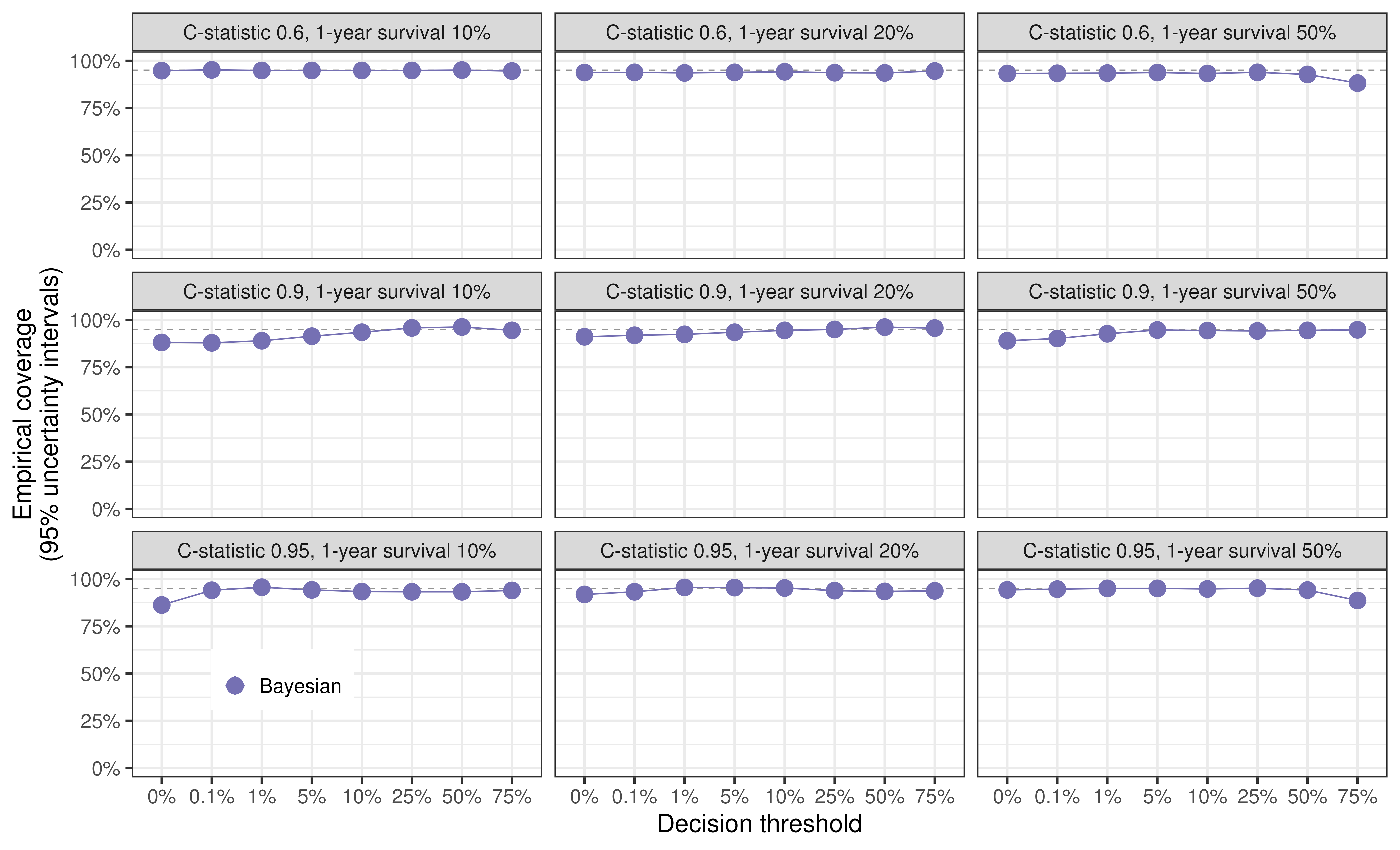}
\end{center}
\caption{\textbf{Bayesian DCA for survival outcomes shows acceptable empirical coverage}. Bayesian DCA was computed using the bayesDCA R package. For each simulation run, DCA was performed for a fixed example model using a simulated test dataset of sample size corresponding to 100 expected events. A total of \simruns Monte Carlo repetitions was run for each setting. The setting C-statistic corresponds to its maximum achievable C-statistic. The example model for each setting was fixed to approximate the maximum discrimination of that setting but was miscalibrated (overly extreme risk predictions). The Bayesian intervals show limited undercoverage and most empirical coverage values are above 90\%.}
\label{fig:empirical-coverage-surv}
\end{figure}

\begin{figure}[H]
\captionsetup{width=.95\linewidth}
\begin{center}
\includegraphics[width=.95\linewidth]{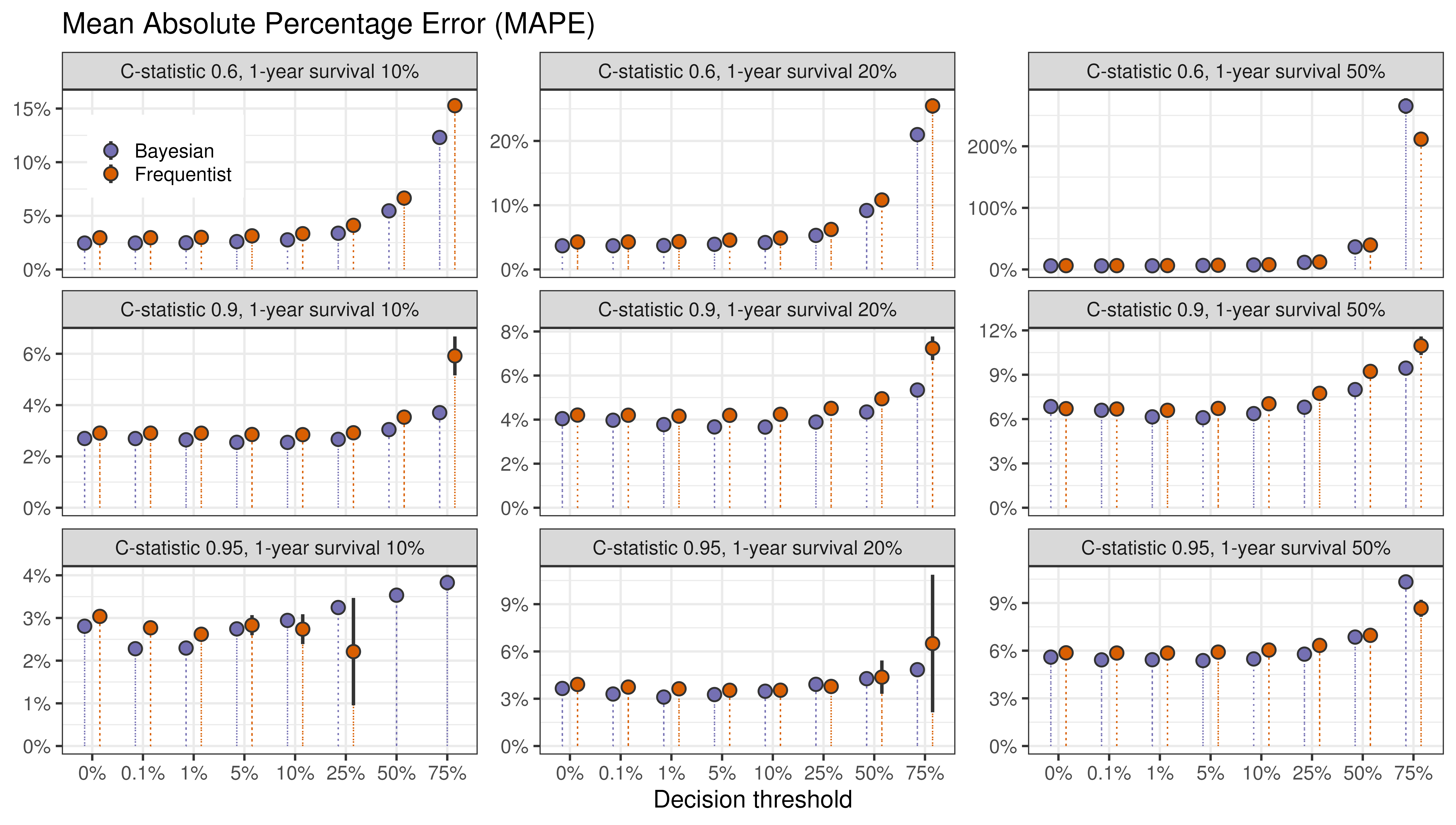}
\end{center}
\caption{\textbf{Bayesian and Frequentist DCA for survival outcomes show similar average estimation errors}. Bayesian DCA was computed using the bayesDCA R package, while the Frequentist alternative used the dcurves package. For each simulation run, DCA was performed for a fixed example model using a simulated test dataset of sample size corresponding to 100 expected events. A total of \simruns Monte Carlo repetitions was run for each setting. The setting C-statistic corresponds to its maximum achievable C-statistic. The example model for each setting was fixed to approximate the maximum discrimination of that setting but was miscalibrated (overly extreme risk predictions). Points show Mean Absolute Percentage Error (MAPE) and ranges show 95\% confidence intervals. In some simulation settings, the Frequentist method fails in the absence of observed survival times past the prediction horizon (e.g., setting with C-statistic 0.95 and 1-year survival 10\%, for thresholds 50\% and 75\%).}
\label{fig:surv-mape}
\end{figure}
\end{document}